\documentclass[modern,trackchanges]{aastex62}

\usepackage{bm} 
\usepackage{amsmath}
\newcommand*\diff{\mathop{}\!\mathrm{d}}
\newcommand{\Secref}[1]{\hyperref[#1]{Section~\ref*{#1}}}
\newcommand{\Appref}[1]{\hyperref[#1]{Appendix~\ref*{#1}}}
\newcommand{\epseri}{$\epsilon$ Eridani}

\graphicspath{{./}{figures/}}

\submitjournal{AJ}

\shorttitle{Bayesian Framework for Exoplanet}
\shortauthors{Ruffio et al.}

\begin{document}

\title{A Bayesian Framework for Exoplanet Direct Detection and Non-Detection}

\correspondingauthor{Jean-Baptiste Ruffio}
\email{jruffio@stanford.edu}
\author[0000-0003-2233-4821]{Jean-Baptiste Ruffio}
\affiliation{Kavli Institute for Particle Astrophysics and Cosmology, Stanford University, Stanford, CA 94305, USA}

\author{Dimitri Mawet}
\affiliation{Department of Astronomy, California Institute of Technology, Pasadena, CA 91125, USA}
\affiliation{Jet Propulsion Laboratory, California Institute of Technology, Pasadena, CA 91109, USA}

\author[0000-0002-1483-8811]{Ian Czekala}
\altaffiliation{KIPAC Postdoctoral Fellow}
\affiliation{Kavli Institute for Particle Astrophysics and Cosmology, Stanford University, Stanford, CA 94305, USA}

\author[0000-0003-1212-7538]{Bruce Macintosh}
\affiliation{Kavli Institute for Particle Astrophysics and Cosmology, Stanford University, Stanford, CA 94305, USA}

\author[0000-0002-4918-0247]{Robert J. De Rosa}
\affiliation{Astronomy Department, University of California, Berkeley; Berkeley CA 94720, USA }

\author{Garreth Ruane}
\altaffiliation{NSF fellow}
\affiliation{Department of Astronomy, California Institute of Technology, Pasadena, CA 91125, USA}

\author{Michael Bottom}
\affiliation{Jet Propulsion Laboratory, California Institute of Technology, Pasadena, CA 91109, USA}

\author{Laurent Pueyo}
\affiliation{Space Telescope Science Institute, Baltimore, MD 21218, USA}

\author[0000-0003-0774-6502]{Jason J. Wang}
\affiliation{Astronomy Department, University of California, Berkeley; Berkeley CA, USA 94720}

\author{Lea Hirsch}
\affiliation{Astronomy Department, University of California, Berkeley; Berkeley CA, USA 94720}

\author{Zhaohuan Zhu}
\affiliation{Department of Physics and Astronomy, University of Nevada, Las Vegas, Las Vegas, NV 89154, USA}

\author[0000-0001-6975-9056]{Eric L. Nielsen}
\affiliation{Kavli Institute for Particle Astrophysics and Cosmology, Stanford University, Stanford, CA 94305, USA}

\begin{abstract}
Rigorously quantifying the information in high contrast imaging data is important for informing follow-up strategies to confirm the substellar nature of a point source, constraining theoretical models of planet-disk interactions, and deriving planet occurrence rates. 
However, within the exoplanet direct imaging community, non-detections have almost exclusively been defined using a frequentist detection threshold (\textit{i.e.} contrast curve) and associated completeness. This can lead to conceptual inconsistencies when included in a Bayesian framework. A Bayesian upper limit is such that the true value of a parameter lies below this limit with a certain probability. The associated probability is the integral of the posterior distribution with the upper limit as the upper bound.
In summary, a frequentist upper limit is a statement about the detectability of planets while a Bayesian upper limit is a statement about the probability of a parameter to lie in an interval given the data. The latter is therefore better suited for rejecting hypotheses or theoretical models based on their predictions.
In this work we emphasize that Bayesian statistics and upper limits are more easily interpreted and typically more constraining than the frequentist approach. We illustrate the use of Bayesian analysis in two different cases: 1) with a known planet location where we also propose to use model comparison to constrain the astrophysical nature of the point source and 2) gap-carving planets in TW Hya. To finish, we also mention the problem of combining radial velocity and direct imaging observations.
\end{abstract}


\section{Introduction}

Direct imaging is a method to spatially resolve exoplanets' light from their host star using large telescopes, adaptive optics, coronagraphs and sophisticated data processing. With ground-based telescopes, this technique currently allows the detection of young ($< 300\,$Myr), massive ($>2\,M_\mathrm{Jup}$), self-luminous exoplanets at host-star separations not yet probed by indirect methods ($a >5\,$AU).
Direct imaging surveys of previously-unobserved stars mostly produce non-detections ($>98\%$ of stars do not have a detectable planet with current instruments). A wide range of science can be drawn from these null results, but thinking about the definition of upper limits is important. We will discuss some of these applications in the following.

First, we consider cases in which the position of the object is known. Most detected point sources are background stars, therefore confirmation of their planetary nature requires follow-up observations \citep{Black1980}. Several strategies can be adopted depending on the information in hand. It is common practice to use the upper limit from the non detection and/or the frequencies of the different astrophysical signals to make the case for a planet and reject the background or foreground hypothesis \citep{Nielsen2017,Chauvin2017,Wagner2016,Macintosh2015,Meshkat2013}. When a point-source has been detected in one of two spectral bands for example, the upper limit on the second band can place limits on its color and in some cases reject the possibility of it being a star. Generally, a red object or one showing significant spectral features characteristic of low-temperature atmospheres will favor lower masses and increase the likelihood of it being bound. This can be a powerful tool to prioritize follow up observations. 
Fomalhaut b is another interesting example of the use of upper limits in determining the nature of an object. The signal was discovered in the optical \citep{Kalas2008} but all subsequent follow-up observations in the infrared yielded non detections casting doubts on its planetary nature \citep{Currie2013,Janson2012,Currie2012,Marengo2009,Kalas2008}. The lack of infrared emission suggests that the signal comes from starlight scattered by a disk surrounding a planetary body.
A more formal statistical approach such as the one derived in this work could be used to set tighter limits on the mass of a self-luminous planet, or to compare different dust formation hypotheses \citep{Kenyon2014}.

Another application of upper limits is to rule out models where an undetected planet perturbs some visible source, \textit{e.g.} by clearing a gap in a circumstellar dust disk. \cite{Ruane2017} used direct imaging data of the TW Hya system to constrain the masses or the accretion rates of hypothetical gap-carving planets. The mass upper limit is derived from the flux constraint using a planet formation model \citep{Baraffe2003,Allard2012}. In this case, the exact position of the planet is not known but the shape of the gap defines its orbit.

A final example of the use of upper limits is combining radial velocity measurements and direct imaging observations. As time baselines keep growing and sensitivity improves, the overlap between the accessible mass and semi-major axis parameter space of the two methods keeps increasing.
When a direct and a radial velocity detection is available, it can be combined to infer the dynamical masses of a binary system. Spectroscopic binary stars that are spatially resolved can be used this way to constrain the age of moving groups \citep{Nielsen2016}. A direct imaging non-detection can still bring useful constraints on the mass of a bound companion \citep{Vaccaro2015,Hardy2015,Joergens2012}.

The use of non-detections in the derivation of exoplanet occurrence rates is also extremely important \citep{Vigan2017,Galicher2016,Bowler2016,Brandt2014,Nielsen2010,Cumming2008}. In this context, combining radial velocity and direct imaging can also help yield better estimates of planet frequency \citep{Lannier2017,Bryan2016}. However, this problem is complex and outside of the scope of this work. Bayesian based occurrences are very sensitive to the accuracy of the noise model, the characterization of which remains an on-going effort for high-contrast imaging.

Planet flux upper-limits are commonly defined using a frequentist approach from a detection threshold (\textit{e.g.}, contrast curve), usually $5\sigma$ with $\sigma$ the standard deviation of the noise \citep{Nielsen2017,Mesa2017,Macintosh2015,Meshkat2013}. In principal, the detection threshold should be derived to set an acceptable false positive rate \citep{Wahhaj2013}, but in practice, is often set to the traditional $5\sigma$ limit. The upper-limit can also be thought of in terms of true positive fraction (\textit{i.e.}, completeness). By definition, the detection threshold corresponds to a $50\%$ completeness. The fundamental conceptual differences between frequentist and Bayesian upper limits have been detailed in the context of gravitational waves detection \citep{Rover2011,Abbott2004,Brady2004,Finn1998}. For example, \cite{Finn1998} emphasizes that Bayesian analysis makes a measure of our degree of belief in a proposition while the Frequentist analysis addresses our confidence in the ability of a procedure to decide if a signal is present or absent, making the Bayesian analysis better suited for the study of individual events. \cite{Rover2011} notes that a frequentist upper limit requires the maximization of the likelihood while the Bayesian upper limits requires integration and also argues that the latter is more easily interpretable.

Although a detection threshold does provide a measure of the depth of the observation, i.e. its sensitivity to faint point sources, it is not a statement about the degree of belief in a given point-source flux given the data. While the detection upper limit indicates which planets would have been detected with a given completeness, a Bayesian upper limit is a statement about the probability of the planet flux given the data.
A Bayesian upper limit $f_{\textrm{lim}}$ is defined from the planet flux posterior and a fixed probability of the true flux to be smaller than this value given the data, $\mathcal{P}(f<f_{\textrm{lim}}|d)$, referred to as the cutoff probability in the following. The cutoff probability is the value of the cumulative distribution of the posterior at the position of the upper limit.
Note that the posterior needs be carefully defined as a function of the question that is asked to the data and the assumptions made.
Using a detection threshold in all circumstances makes the interpretation of the results more difficult.
The existing examples of combining radial velocity and direct imaging measurements treat the radial velocity data in a Bayesian framework while using the frequentist approach for direct imaging upper limits. Using the concepts presented here, Mawet et al. (2018, submitted to AJ) will be a step toward treating both data types in a consistent Bayesian framework.

We will illustrate this approach for direct imaging by revisiting practical cases of non-detection.
\Secref{sec:knownPos} assumes the location of the planet known and proposes to use Bayesian model comparison to decide the most probable nature of a candidate.
\Secref{sec:knownOrbit} considers the case of a known orbit for the companion but no information on its precise location. This case is applicable to constraining the mass of an undetected accreting planet in a disk gap for example.
In \Secref{sec:RVIM}, we look at combining radial velocity and direct imaging measurements to constrain planet mass and orbital parameters.
We conclude in \Secref{sec:conclusion}.

\section{Companion at a known location}
\label{sec:knownPos}

\subsection{Bayes' rule and upper limit}
\label{sec:upperlimit}

In this section, we address the simple problem of defining an upper limit for the flux of a point source at a known location. A typical example is when the planet was clearly detected in only one of two spectral bands. In the literature, quoted upper limits are often defined as the detection threshold (commonly $5\sigma$) at the location of the object.

\begin{figure}
\centering
\includegraphics[width=\linewidth]{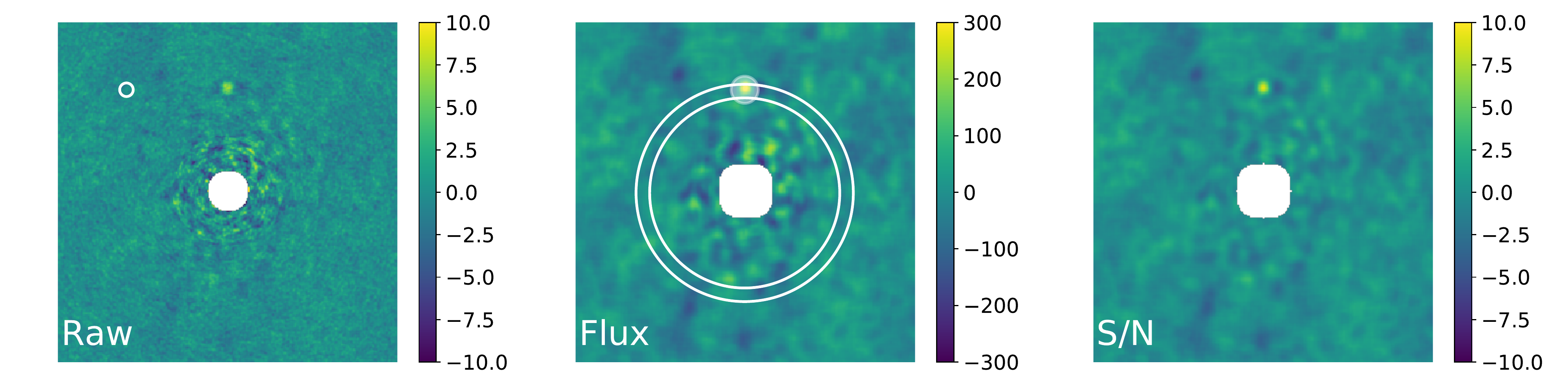}
\caption{Processed images of a KeckII/NIRC2 observation of TW Hya \citep{Ruane2017} including a simulated $10\sigma$ point source located North of the center star. The leftmost image is the combined dataset after the speckles were removed (The units are arbitrary). The middle image is a flux map resulting from cross correlating  a 10 pixel wide aperture (white empty circle in the left image) across the image. For each position, the sample standard deviation is computed from a $10$ pixel-wide annulus at the same separation after the surroundings of the pixel of interest have been masked out (white filled circle). The rightmost image is the resulting Signal to Noise ratio (S/N) map.}
\label{fig:classicalKLIP}
\end{figure}

\autoref{fig:classicalKLIP} shows a typical example of a high contrast image\footnote{Observation from \cite{Ruane2017} in L' ($3.4-4.1\mu m$) including 357 single exposures totaling $4.5$ hours of integration time. } in which a simulated $\approx10\sigma$ point source was injected North of the center star. The left most image is the result of combining the single exposures after the speckle noise was removed using a principal component based approach \citep{Wang2015,Soummer2012}. The simulated point source was injected in the individual 357 frames before the speckle subtraction is performed. The simplest approach to compute the flux of a point source is to use aperture photometry, which consist in integrating the flux inside an aperture with a diameter equal to the width of the point spread function (PSF) \citep{Mawet2014}. Such a flux map is equivalent to the cross correlation\footnote{For example using the Python function scipy.signal.correlate2d.} of that aperture with the image (middle panel in \autoref{fig:classicalKLIP}). A more accurate flux could be estimated using a matched filter \citep{Ruffio2017} as shown in \Secref{sec:knownOrbitdef}. For each position, the error bar $\sigma$ on the flux estimate can be defined as the sample standard deviation computed from a $10$ pixel-wide annulus at the same separation from the primary star after the surroundings of the pixel of interest have been masked out (white filled circle) \citep{Mawet2014,Marois2008}. This is the same standard deviation that is also used to defined the detection threshold as a function of separation. A Signal to Noise ratio (S/N) map is simply the flux map divided by the standard deviation at each separation (Right most image in \autoref{fig:classicalKLIP}). Any signal brighter than $5\sigma$ is generally flagged as a possible candidate. The threshold can be defined such as to yield a reasonable number of false positives over a fixed field of view. In a direct imaging survey, it can be limited by the number of candidates on which follow-up observations can be performed or more generally by the follow-up strategy that maximizes the science return of the survey.

\begin{figure}[bth]
\gridline{\fig{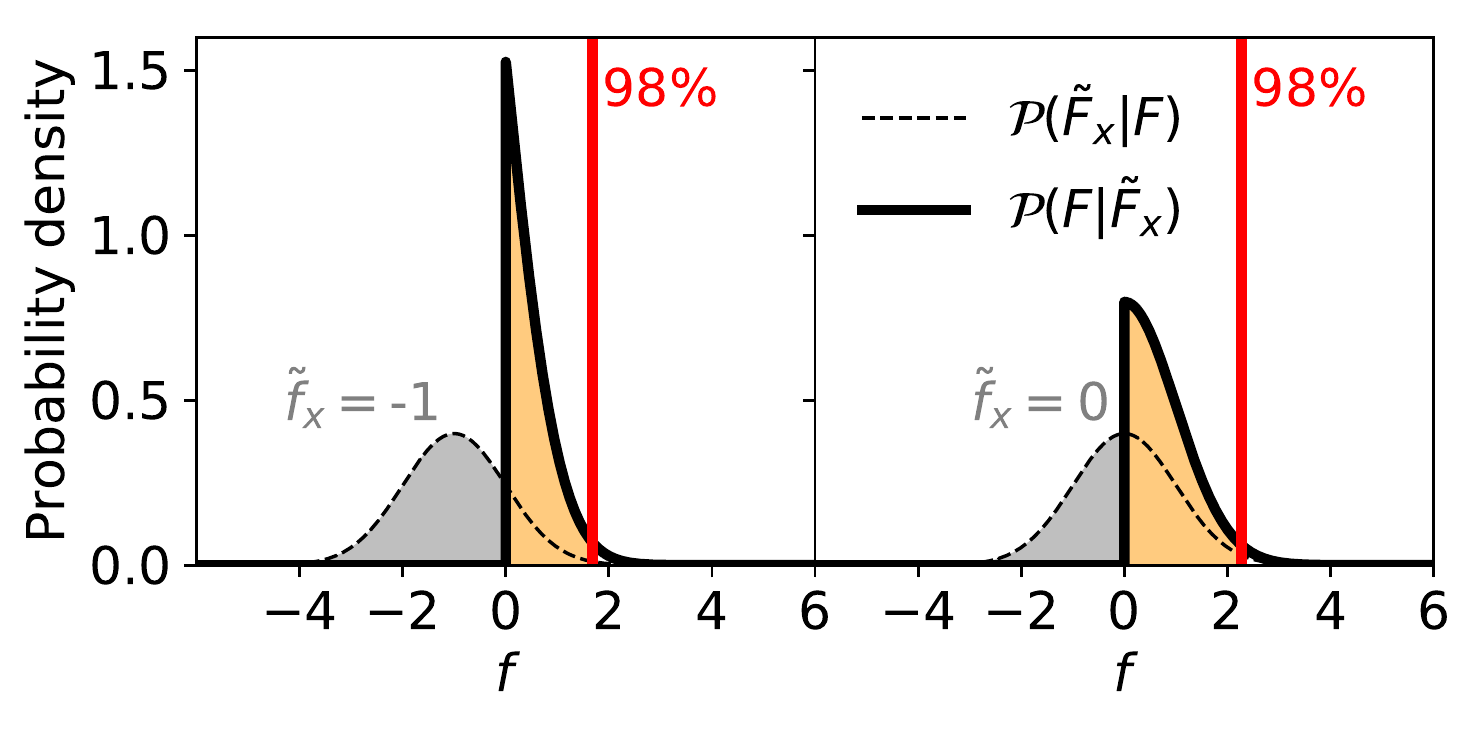}{0.5\textwidth}{}
		\fig{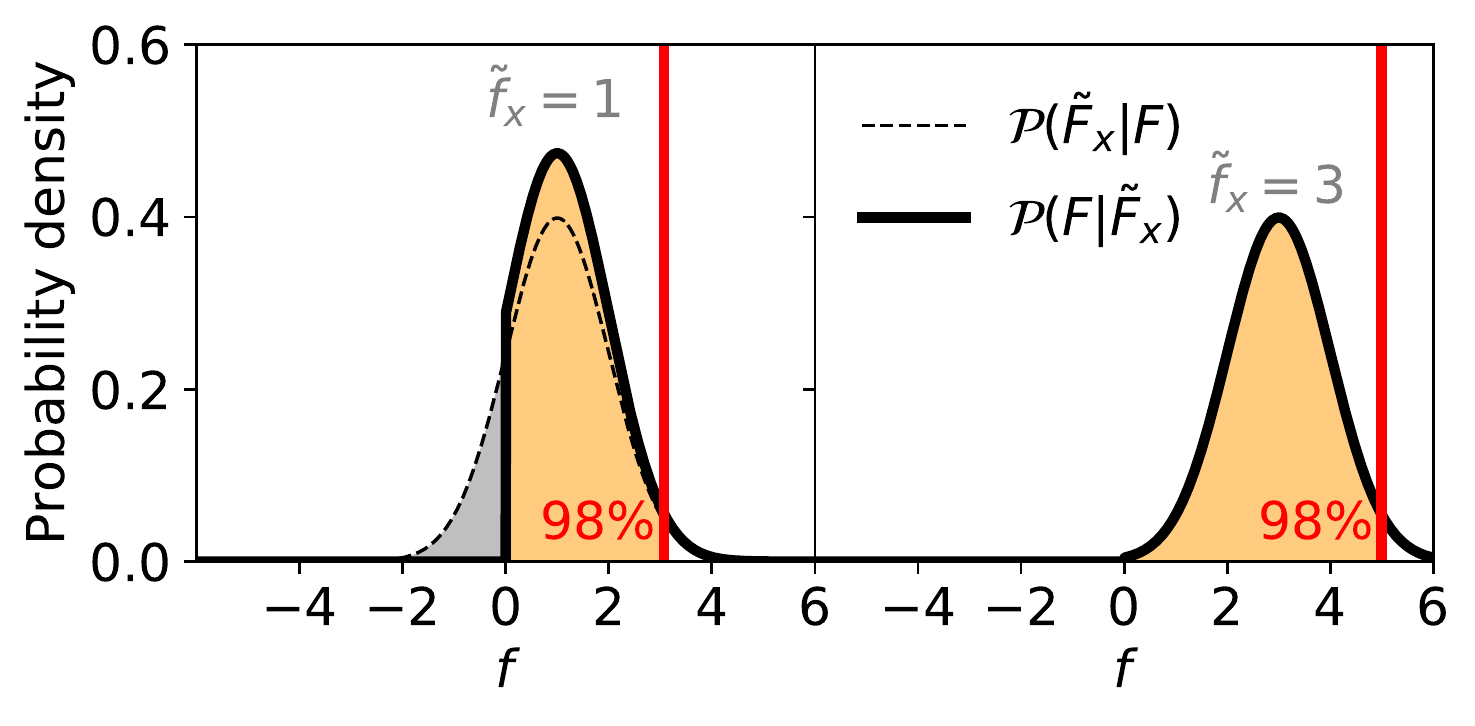}{0.5\textwidth}{}
          }
\caption{flux posterior (solid line) and upper-limit as a function the estimated flux. The dashed line is the likelihood centered on the estimated flux, for which we assume a unit standard deviation. We assume a different S/N for each plot ($-1\sigma$,$0\sigma$,$1\sigma$ and $3\sigma$). The gray area represents the values of the flux that are rejected by the positive prior. The upper limit is defined from a $97.7\%$ cutoff probability represented by the orange area under the curve. The different y-scales are due to the necessary normalization of the posteriors. For the $3\sigma$ case, the likelihood is hidden in the width of the line from the posterior because the effect of the prior is negligible. }
\label{fig:posteriorKnownPos}
\end{figure}

Such a detection threshold is fundamentally not a statement about the parameter space that was ruled out by the data.
An upper-limit is better understood in the context of Bayesian inference. One first needs to define a probability, hereafter cutoff probability, of the true planet flux to fall below the upper limit. The Bayesian upper limit is then defined as the value for which the cumulative distribution of the posterior is equal to the cutoff probability.
From now on, we will make a distinction between such an upper limit and a detection threshold.
For a Gaussian noise with known standard deviation $\sigma$, an unbounded uniform prior and a $97.7\%$ cutoff probability, the upper limit is $2\sigma$ above the estimated flux (See \autoref{fig:posteriorKnownPos}). The flux can be estimated even if the planet is not formally detected as long as its position is known. In the unlucky event of a very negative noise sample (for example lower than $-2\sigma$) at the location of the planet, the estimated flux and the upper-limit could become negative. This might be unsettling as we know that a flux is strictly positive, but this will be corrected by a more informative prior, which will forbid negative values of the flux. For a given noise distribution, it is important to note that the upper-limit is a function of the data, here the estimated flux, while the detection threshold is a property of the noise. As we said, the choice of the threshold is also somewhat arbitrary, whereas a posterior is entirely defined by the properties of the noise, the data and the choice of a prior. As a consequence, using the detection threshold in place of an upper limit is not making optimal use of the data.

Let $F$ be the true planet flux random variable and $\tilde{F}_x$ as its estimate at the position $x$ based on the observation. We denote random variables and random vectors with an upper case and their realization with a lower case.
The posterior $\mathcal{P}(F \vert \tilde{F}_x)$ is the probability of the point-source flux given its estimated value from the observation. In this context, we define the data as being the flux map and following Bayes' rule \citep{SiviaBook},
\begin{equation}
\mathcal{P}(F \vert \tilde{F}_x) = \frac{ \mathcal{P}(\tilde{F}_x \vert F) \mathcal{P}(F)}{ \mathcal{P}(\tilde{F}_x)},
\label{eq:bayesrule}
\end{equation}
where $\mathcal{P}(\tilde{F}_x \vert F)$ is the likelihood, $\mathcal{P}(F)$ the flux prior and $\mathcal{P}(\tilde{F}_x)$ acts as a normalization factor to ensure the posterior distribution has a unit integral.
If the noise is Gaussian, the likelihood $\mathcal{P}(\tilde{F}_x \vert F)$ is given by the Gaussian distribution,
\begin{equation}
\mathcal{P}(\tilde{F}_x =\tilde{f}_x \vert F=f) = \frac{1}{\sqrt{2\pi}\sigma} \exp\left\lbrace-\frac{1}{2}\frac{(f-\tilde{f}_x)^2}{\sigma^2}\right\rbrace .
\label{eq:gausslike}
\end{equation}

The term $\mathcal{P}(\tilde{F}_x)$ is also called the marginal likelihood (or model evidence) and can be written as,
\begin{equation}
\mathcal{P}(\tilde{F}_x) = \int_{f=-\infty}^{\infty} \mathcal{P}(\tilde{F}_x \vert F=f) \mathcal{P}(F=f) \diff f.
\end{equation}
The marginal likelihood is used in Bayesian model comparison.
In the following, we use a simple uniform positive prior,
\begin{equation}
\mathcal{P}(F=f) = \begin{cases} \alpha, & \mbox{if } f > 0\mbox{, with } \alpha \mbox{ constant} \\ 0, & \mbox{if } f \le 0 \end{cases} .
\end{equation}
The uniform prior corresponds to the objective Jeffreys' prior (\textit{i.e.}, square root of the determinant of the Fisher information) for location parameters like the mean of a normal distribution \citep{KassWasserman1996}. The uniform prior is improper (\textit{i.e.}, its integral is infinite) but the likelihood will be very constraining for large values of the flux $f$ so an upper bound for the prior of the companion flux will not make a significant difference.
The inverse prior $\mathcal{P}(f) \propto 1/f$ is a common choice for positive real valued parameters because it is the only prior invariant by rescaling, \textit{i.e.}, $\mathcal{P}(f)\diff f = \mathcal{P}(\beta f)\diff(\beta f)$. It means that the inverse prior does not favor any particular scale of the parameter, but it requires a strictly positive lower bound to produce a normalizable posterior. The problem is that the previous likelihood does not constrain null values, which means that the inverse prior will be dominating at small values of the flux. As a consequence, due to the divergence of the integral of the inverse function, the upper limit will be extremely dependent on the choice of the lower bound in the prior. For example, the upper limit would tend to zero for an infinitely small lower bound. The prior could also be derived from our current knowledge of planet population. This would suggest a log-uniform planet mass prior \cite{Cumming2008} and the corresponding flux prior after a change of variable, but this would lead to the same difficulty near zero.
In this context, the uniform prior remains a conservative choice for the definition of an upper limit. Indeed, a more relevant prior based on a planet population model would give more weights to lower fluxes and therefore decrease the upper limit. The uniform prior also facilitates the comparison of upper limits resulting from different works, because it does not include a user-defined parameter. Note that in the absence of a prior, the posterior is simply equal to the likelihood centered on the estimated flux. 

We will assume Gaussian noise in the following, but the previous statements apply irrespective of the choice of noise distribution.
We now define the cumulative distribution of a Gaussian distribution with standard deviation $\sigma$ and mean $\tilde{f}$ as
\begin{equation}
\mathcal{C}_{\tilde{f},\sigma}(f_0)=\int_{f=-\infty}^{f_0}\frac{1}{\sqrt{2\pi}\sigma} \exp\left\lbrace-\frac{1}{2}\frac{(f-\tilde{f})^2}{\sigma^2}\right\rbrace \diff f.
\end{equation}
and its inverse $\mathcal{Q}$, also known as the quantile function, such that $f_0 = \mathcal{Q}(\mathcal{C}(f_0))$.

The flux upper limit $f_{\textrm{lim}}$ is the value for which the cumulative distribution of the posterior is equal to a cutoff probability $\eta$ (for example $97.7\%$), \textit{i.e.},
\begin{equation}
\eta =\int_{f=-\infty}^{f_{\textrm{lim}}} \mathcal{P}(f \vert \tilde{f}_x) \diff f = \int_{f=-\infty}^{f_{\textrm{lim}}} \frac{ \mathcal{P}(\tilde{f}_x \vert f) \mathcal{P}(f)}{ \mathcal{P}(\tilde{f}_x)} \diff f,
\label{eq:cumuldef}
\end{equation}
with $\tilde{f}_x$ and $\sigma_x$ the estimated flux and standard deviation at the position $x$. We also used \autoref{eq:bayesrule} in the second equality.

Then, we note that using Gaussian statistics $\mathcal{P}(\tilde{f}_x) = 1-\mathcal{C}_{\tilde{f}_x,\sigma_x}(0)$ and using a positive uniform prior $\int_{f=-\infty}^{f_{\textrm{lim}}} \mathcal{P}(\tilde{f}_x \vert f) \mathcal{P}(f) \diff f = \mathcal{C}_{\tilde{f}_x,\sigma_x}(f_{\textrm{lim}} )-\mathcal{C}_{\tilde{f}_x,\sigma_x}(0)$, which leads to  
\begin{equation}
\eta = \frac{\mathcal{C}_{\tilde{f}_x,\sigma_x}(f_{\textrm{lim}} )-\mathcal{C}_{\tilde{f}_x,\sigma_x}(0)}{1-\mathcal{C}_{\tilde{f}_x,\sigma_x}(0)}.
\end{equation}
After rearranging the terms and using the quantile function, one finds
\begin{equation}
f_{\textrm{lim}} = \mathcal{Q}_{\tilde{f}_x,\sigma_x}\left(\eta+(1-\eta)\mathcal{C}_{\tilde{f}_x,\sigma_x}\left(0\right)\right).
\label{eq:upperlimit}
\end{equation}

\autoref{fig:posteriorKnownPos} illustrates the posterior and upper-limit for different measured fluxes assuming a unit standard deviation. Note that if we drop the positivity constraint on the prior $\mathcal{P}(F)$, or if the estimated flux $\tilde{f}_x$ is large, the $97.7\%$ upper-limit does correspond to $2\sigma$ above the measured value. \autoref{fig:posteriorKnownPos} highlights how the upper limit closely depends on the realization of the noise, which is not true of the detection threshold. We encourage the direct imaging community to consider quoting upper-limits based on this definition instead of a detection threshold.

We have assumed that the standard deviation of the flux estimate was known. In practice, the standard deviation is estimated from the data, which means that it needs to be marginalized over when it is poorly constrained (e.g., when the number of noise realizations in the annulus of \autoref{fig:classicalKLIP} is small). The planet flux posterior can be defined as the two-sample t-test in which one of the samples has one element (corresponding to the location of the planet) and the other sample contains all the pixels of a region with similar noise properties but free of astrophysical signal \citep{Mawet2014}. The second sample is often defined as the pixels taken at the same projected separation but located one resolution element apart from each other. The goal of a two-sample t-test, with unequal sample sizes but equal variance, is to estimate the difference between the means of two samples given that the variance is unknown and must be estimated from the data itself. In this case, the differences of the means is no other that the planet flux and its posterior must be marginalized over the uncertainty of the sample means and standard deviation, which result in a Student-t distribution: 
\begin{equation}
\mathcal{P}(F=f \vert D) \propto \left[\frac{t_f}{N-1}+1\right]^{-((N-1)+1)/2}, \label{eq:studenttposterior}
\end{equation}
with
\begin{equation}
t_f = \frac{(f-\tilde{f}_x)^2}{(\sum_k \tilde{f}_k^2/(N-1))\sqrt{1/N+1}},
\end{equation}
where $N$ is the number of elements without astrophysical signal.
\autoref{eq:upperlimit} can still be used after redefining $\mathcal{C}$ and $\mathcal{Q}$ using \autoref{eq:studenttposterior} in place of the normal distribution.

\subsection{Bayesian model comparison: is it a star?}

A recurring problem in direct imaging is to constrain the astrophysical nature of a candidate given a set of concurrent observations in different spectral bands. In this context, the nature of a point-like source can for example be a background star ($\mathcal{H}_\star$), a galaxy, a brown-dwarf (background/foreground or gravitationally bound), a planet or a false positive ($\emptyset$), also-known-as null hypothesis.
Point-source detections are a common occurrence in direct imaging surveys due to the prevalence of background stars, which is why we would like to prioritize their follow-up observations to optimize the discovery of new planets. 
For the purpose of this section, we will assume that we have two broad-band observations, one of which could be a non detection. With a single detection, it is not possible to compare the astrometric measurements of the candidate to that of a background object. 
It is common practice to use the upper limit from the non detection and/or the frequencies of the different astrophysical signals to make the case for a planet and reject the background or foreground hypothesis \citep{Nielsen2017,Chauvin2017,Wagner2016,Macintosh2015,Meshkat2013}. 

Deciding between alternative hypotheses, which are here the possible classes of astrophysical signal, can be done more generally using Bayesian model comparison.
Given an hypothesis $\mathcal{H}$, data $D$ and model parameters $\Theta$, the marginal probability $\mathcal{P}(D|\mathcal{H})$ is the probability of obtaining the data assuming that an hypothesis is true. It is also defined as the normalization factor in the denominator of Bayes' rule:
\begin{equation}
\mathcal{P}(\Theta|D,\mathcal{H}) =\frac{\mathcal{P}(D|\Theta,\mathcal{H}) \mathcal{P}(\Theta|\mathcal{H}) }{\mathcal{P}(D|\mathcal{H}) }.
\end{equation}
This marginal probability also appears in the expression of the posterior probability of an hypothesis $\mathcal{P}(\mathcal{H}|D) \propto \mathcal{P}(D|\mathcal{H}) \mathcal{P}(\mathcal{H})$. The ratio of the posterior probability of two hypotheses $\mathcal{H}_1$ and $\mathcal{H}_2$, called the Bayes factor, is then given by
\begin{equation}
\frac{\mathcal{P}(\mathcal{H}_2|D)}{\mathcal{P}(\mathcal{H}_1|D)} = \frac{\mathcal{P}(D|\mathcal{H}_2)\mathcal{P}(\mathcal{H}_2)}{\mathcal{P}(D|\mathcal{H}_1)\mathcal{P}(\mathcal{H}_1)},
\end{equation}
with $\mathcal{P}(\mathcal{H}_1)$ and $\mathcal{P}(\mathcal{H}_2)$ the prior probability of each hypothesis to be true. The hypothesis $\mathcal{H}_2$ is preferred when the ratio is large.
The significance of a given Bayes factor can be read out of published tables \citep[see \autoref{tab:Bayesfactor},][]{JeffreysBook,Kass1995,Robert15112}.

\begin{deluxetable}{ccc}
\tablecaption{ Evidence against $\mathcal{H}_1$ compared to $\mathcal{H}_2$ given the value of the Bayes factor, $B=\mathcal{P}(\mathcal{H}_2|D)/\mathcal{P}(\mathcal{H}_1|D)$, from \cite{Kass1995}.}
\tablecolumns{3}
\tablewidth{0pt}
\tablehead{
\colhead{$\log_{10}(B)$} &
\colhead{$B$} &
\colhead{Evidence against $\mathcal{H}_1$}
}
\startdata
\label{tab:Bayesfactor}
$0$ to $1/2$ & $1$ to $3.2$ & Not worth more than a bare mention\\
$1/2$ to $1$ & $3.2$ to $10$ & Substantial \\
$1$ to $2$ & $10$ to $100$ & Strong \\
$>2$ & $>100$ & Decisive \\
\enddata
\end{deluxetable}

In the same way as we defined $\tilde{F}_x$, we write $\tilde{G}_x$ as the estimated photometry in a second spectral band. 
The probability of the null hypothesis $\mathcal{H}=\emptyset$ given the observations, $\tilde{F}_x$ and $\tilde{G}_x$, is given by
\begin{equation}
\mathcal{P}(\emptyset \vert \tilde{F}_x,\tilde{G}_x) = \frac{\mathcal{P}(\tilde{F}_x,\tilde{G}_x \vert \emptyset) \mathcal{P}(\emptyset)}{\mathcal{P}(\tilde{F}_x,\tilde{G}_x)},
\end{equation}
with  $\mathcal{P}(\emptyset)$ the prior probability of not having an astrophysical signal. Given that the two observations are independent (\textit{i.e.}, $\mathcal{P}(\tilde{F}_x,\tilde{G}_x\vert \emptyset)=\mathcal{P}(\tilde{F}_x \vert \emptyset)\mathcal{P}(\tilde{G}_x \vert \emptyset)$) and assuming Gaussian distributions, the likelihood given the null hypothesis is defined as
\begin{equation}
\mathcal{P}(\tilde{F}_x=\tilde{f}_x,\tilde{G}_x=\tilde{g}_x \vert \emptyset) = \frac{1}{2\pi\sigma_{fx}\sigma_{gx}} \exp\left\lbrace-\frac{1}{2}\frac{\tilde{f}_x^2}{\sigma_{fx}^2}-\frac{1}{2}\frac{\tilde{g}_x^2}{\sigma_{gx}^2}\right\rbrace .
\end{equation}

We define $m_1$ and $m_2$ as the magnitude in each spectral band and $m_{1/2}=m_1-m_2$ as the color. If $\mathcal{H}$ represents a given hypothesis of the astrophysical nature of a point source, then
\begin{align}
\mathcal{P}(\mathcal{H} \vert \tilde{F}_x,\tilde{G}_x) &= \int_{m_1,m_{1/2}} \mathcal{P}(\mathcal{H},m_1,m_{1/2} \vert  \tilde{F}_x,\tilde{G}_x) \diff m_1 \diff m_{1/2} ,\nonumber \\
&= \int_{m_1,m_{1/2}} \mathcal{P}(\tilde{F}_x,\tilde{G}_x \vert \mathcal{H},m_1,m_{1/2}) \mathcal{P}(m_1,m_{1/2} \vert \mathcal{H})  \diff m_1 \diff m_{1/2} \ \ \frac{ \mathcal{P}(\mathcal{H})}{\mathcal{P}(\tilde{F}_x,\tilde{G}_x)} . \nonumber \\
\label{eq:starproba}
\end{align}
From right to left, $\mathcal{P}(\mathcal{H})$ is prior probability of the hypothesis, $\mathcal{P}(m_1,m_{1/2} \vert \star)$ is the prior probability of a the color-magnitude of an object defined by $\mathcal{H}$, $\mathcal{P}(\tilde{F}_x,\tilde{G}_x \vert \star,m_1,m_{1/2})$ is the likelihood, and to finish, $\mathcal{P}(\tilde{F}_x,\tilde{G}_x)$ is the marginal likelihood. $\mathcal{P}(\mathcal{H})$ is therefore defined as the frequency of objects $\mathcal{H}$ in a small arbitrary box at the position $x$ on the detector and $\mathcal{P}(m_1,m_{1/2} \vert \star)$ is the color-magnitude distribution of these objects. The likelihood is given by
\begin{equation}
\mathcal{P}(\tilde{F}_x=\tilde{f}_x,\tilde{G}_x=\tilde{g}_x \vert \mathcal{H},m_1,m_{1/2}) = \frac{1}{2\pi\sigma_{fx}\sigma_{gx}} \exp\left\lbrace-\frac{1}{2}\frac{(\tilde{f}_x-f)^2}{\sigma_{fx}^2}-\frac{1}{2}\frac{(\tilde{g}_x-g)^2}{\sigma_{gx}^2}\right\rbrace ,
\end{equation}
with $f = 10^{-m_1/2.5}$ and $g = 10^{-m_2/2.5}$. $\sigma_{fx}$ and $\sigma_{gx}$ are the error bars respectively for the flux estimates $\tilde{f}_x$ and $\tilde{g}_x$.

In order to carry on our hypothetical example, we assume that a first observation was made with the $4.4 \mu m$ filter (F444W) of the NIRCAM instrument on-board the James Webb Space Telescope (JWST). The follow-up observation is done in H band with the Gemini Planet Imager (GPI). The goal is to identify the most likely nature of a candidate given the JWST detection and the GPI data. For the sake of simplicity, we will only consider two stellar populations (high mass noted $\mathcal{H}_{\star,\mathrm{high}}$, low mass noted $\mathcal{H}_{\star,\mathrm{low}}$) and the null hypothesis to illustrate the classification method. However, we would like to emphasize that this framework is in no way restricted to this example and should in practice at least include models of planets or brown-dwarfs.

\begin{figure}
\centering
\includegraphics[width=0.5\linewidth]{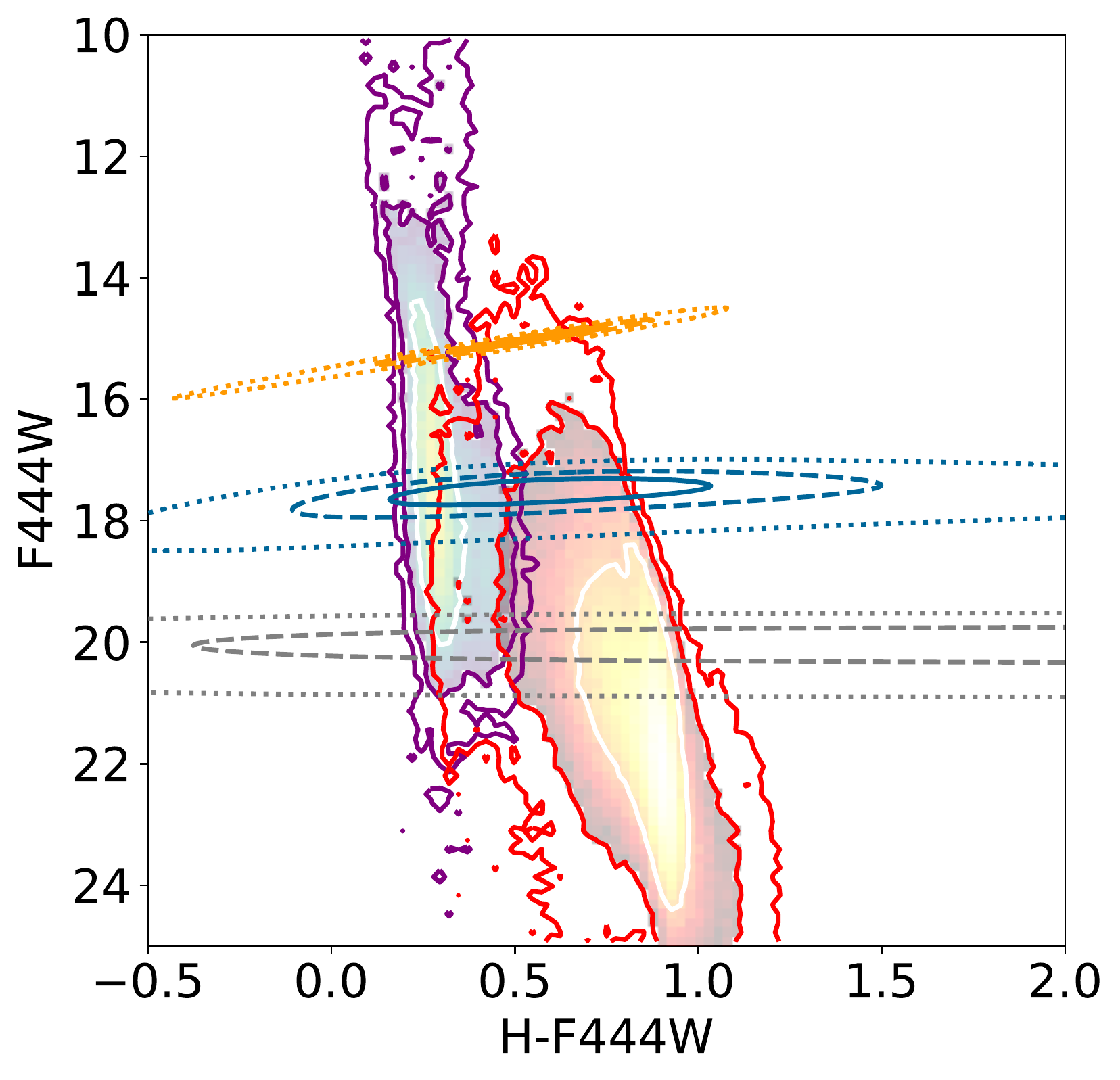}
\caption{Color and apparent magnitude diagram of stars compared to observations. The colormaps correspond to the density of low-mass (red) and high-mass stars ($>0.65M_{\odot}$, purple), which is only drawn up to a $95\%$ confidence levels. The inner (outer) contour represents the $68\%$ ($99.7\%$) confidence level contours. The elongated contours represent the $68\%$ (solid), $95\%$ (dashed) and $99.7\%$ (dotted) confidence levels of the likelihood for the three scenarios described in \autoref{tab:JWSTGPI}. The probabilities of each candidate to be respectively a low-mass or high-mass star is shown in \autoref{tab:JWSTGPIresults}. }
\label{fig:starproba}
\end{figure}

The prior probability distributions of finding a background star as function of their position in a color-magnitude diagram, $\mathcal{P}(m_1,m_{1/2} \vert \mathcal{H}_{\star,\mathrm{high}})$ and $\mathcal{P}(m_1,m_{1/2} \vert \mathcal{H}_{\star,\mathrm{low}})$, can be calculated from the Besan\c{c}on model of stellar populations \citep{Robin2003}.  
We generated a galactic population model\footnote{Using \url{http://model.obs-besancon.fr} } within a solid angle of $0.23\ \textrm{deg}^{2}$ in the vicinity of \object{p Puppis} (\object{HD 60863}, $l=242.96^{\circ}$, $b=-3.87^{\circ}$). The area was chosen to generate approximately $2\times 10^5$ stars within 125\,kpc and with an apparent magnitude of $K<28$. White dwarfs were filtered out of the catalog based on their surface gravity. Magnitudes for each star within the simulation were calculated using a grid of stellar atmospheres \citep{Castelli2004} and the properties of each star reported in the catalog; the distance, effective temperature, surface gravity, metallicity, radius, and extinction (using $R_V = 3.1$, $A_H/A_V=0.184$, and $A_{4.4}/A_V=0.0)$. \autoref{fig:starproba} shows the color-magnitude diagram of the catalog in the JWST/F444W and GPI/H filters. We identify two families of stars, low and high mass (198,027 and 32,292 stars), with a boundary at $0.65M_{\odot}$, which corresponds to the soft boundary between the two modes of the two-dimensional histogram. 
A low-mass star explanation will be preferred when the ratio $\mathcal{P}(\mathcal{H}_{\star,\mathrm{low}} \vert \tilde{F}_x,\tilde{G}_x)/\mathcal{P}(\mathcal{H}_{\star,\mathrm{high}} \vert \tilde{F}_x,\tilde{G}_x)$ is large. The priors $\mathcal{P}(m_1,m_{1/2} \vert \mathcal{H}_{\star,\mathrm{high}})$ and $\mathcal{P}(m_1,m_{1/2} \vert \mathcal{H}_{\star,\mathrm{low}})$ are respectively the normalized two-dimensional histograms in color-magnitude resulting from the stellar population model.

The prior probabilities $\mathcal{P}(\mathcal{H}_{\star,\mathrm{high}})$ and $\mathcal{P}(\mathcal{H}_{\star,\mathrm{low}})$ are defined as the frequency of such stars in an patch of the sky corresponding to a GPI resolution element (a $\approx50$ mas diameter circle). The probability of the null hypothesis is defined such that $\mathcal{P}(\emptyset) + \mathcal{P}(\mathcal{H}_{\star,\mathrm{high}}) + \mathcal{P}(\mathcal{H}_{\star,\mathrm{low}})=1$. The size of the patch of sky will not impact the ratios of probabilities of low-mass and high-mass stars but it will influence the relative probability of the null hypothesis, which is why the probability of the null hypothesis should not be confused with a false positive rate. 

The likelihood is a function of the fluxes and associated error bars of the candidate in the different spectral bands. Assuming that some sources were detected in JWST at $5\sigma$ and subsequently observed by GPI, we describe three possible scenarios corresponding to three candidates. In these scenarios, we vary the sensitivity of the JWST observation and the S/N of the GPI follow up assuming a clear detection ($20\sigma$) and two non detections ($3\sigma$,$-1\sigma$). 
The contours of the likelihood for each of the candidates are drawn in \autoref{fig:starproba}, which is therefore an illustration of the integrand of \autoref{eq:starproba}, $\mathcal{P}(\tilde{F}_x,\tilde{G}_x \vert \mathcal{H}_{\star},m_1,m_{1/2}) \mathcal{P}(m_1,m_{1/2} \vert \mathcal{H}_{\star})$.
The details of the parameters are described in \autoref{tab:JWSTGPI} and the results are presented in \autoref{tab:JWSTGPIresults}.

We can conclude that the first candidate (orange) is most likely a high mass star. The probability of a low mass star is still high because low-mass star are five times more common that high-mass stars in this catalog. The nature of the second candidate is undecided because the probabilities are too similar. However it is still unlikely to be a false positive. 
The follow-up epoch of the third candidate would be classified as a non-detection and yielded a negative flux, which means that it is entirely dominated by the noise. However, stars are still located in the $2\sigma$ region of the likelihood. The first epoch $5\sigma$ detection still makes it somewhat unlikely for the candidate to be pure noise, and the low-mass star hypothesis is preferred.

\begin{deluxetable}{ccccccc}
\tablecaption{$5\sigma$ sensitivity in apparent magnitude, S/N and apparent magnitude of the point source for three hypothetical candidates detected with JWST/NIRCAM with the F444W $4.4 \mu m$ filter and followed-up with GPI in H band. These candidates are compared to a population of stars in \autoref{fig:starproba}. }
\tablecolumns{7}
\tablewidth{0pt}
\tablehead{
 & \multicolumn{3}{c}{F444W (NIRCAM)} & \multicolumn{3}{c}{H (GPI)}\\
\colhead{Candidate} &
\colhead{$5\sigma$} &
\colhead{S/N} &
\colhead{Mag} &
\colhead{$5\sigma$} &
\colhead{S/N} &
\colhead{Mag}
}
\startdata
\label{tab:JWSTGPI}
1 (Orange) & 15 & 5 & 15 & 17.5 & 30 & 15.6 \\
2 (Blue) & 17.5 & 5 & 17.5 & 17.5 & 3 & 18.1 \\
3 (Grey) & 20 & 5 & 20 & 17.5 & -1 & N/A \\
\enddata
\end{deluxetable}

\begin{deluxetable}{ccccc}
\tablecaption{Model probabilities for the three scenarios described in \autoref{tab:JWSTGPI} and illustrated in \autoref{fig:starproba}.}
\tablecolumns{5}
\tablewidth{0pt}
\tablehead{
\colhead{Candidate} &
\colhead{$\mathcal{H}_{\star,\mathrm{high}}$} &
\colhead{$\mathcal{H}_{\star,\mathrm{low}}$} &
\colhead{$\emptyset$} &
\colhead{$\mathcal{H}_{\star,\mathrm{low}}/\mathcal{H}_{\star,\mathrm{high}}$}
}
\startdata
\label{tab:JWSTGPIresults}
$\mathcal{P}(\mathcal{H})$ prior\tablenotemark{a} & $10^{-4}$ & $5\times10^{-4}$ & $0.9994$ & 5\\
1 (orange) & 0.81 & 0.19 & 0.0 & 0.20 \\
2 (Blue) & 0.50 & 0.50 & 0.002 & 0.84 \\
3 (Grey) & 0.04 & 0.88 & 0.08 & 16.2 \\
\enddata
\tablenotetext{a}{The priors $\mathcal{P}(\mathcal{H}_{\star,\mathrm{high}})$ and $\mathcal{P}(\mathcal{H}_{\star,\mathrm{low}})$ are defined from the frequency of the corresponding kind of stars from the Besan\c{c}on stellar population model \citep{Robin2003}.}
\end{deluxetable}

\section{Companion on a known orbit}
\label{sec:knownOrbit}

\subsection{Definition}
\label{sec:knownOrbitdef}

In this section, we will assume that we know the orbit of a planet projected onto the sky plane, but that we do not know its position along it. For example, this situation arise when trying to constrain the mass of a planet in the gap of a proto-planetary disk \citep{Ruane2017}, where the geometry of the gap defines its orbit \citep{Dong2017}. Future astrometric discoveries of planets could also provide orbits of unseen planets \citep{Perryman2014}.

We define the data $D$, or observation, as the random vector representing the pixel values of the image. The point-source parameters are its position on the projected orbit defined as the curvilinear abscissa $S$ and its flux $F$.
We also define $N$ as a Gaussian random vector with zero mean and covariance matrix $\Sigma$. In practice, the noise is assumed to be independent in which case $\Sigma$ becomes diagonal. Data, signal and noise are related through,
\begin{equation}
D=Fm+N,
\end{equation}
with $m=m(s)$ the planet model in the direct imaging data, which is effectively a function of the planet position $S$. We assume that $m$ (i.e. the shape of the PSF) is independent of the flux $F$. When the planet is located outside the field of view or inside the focal plane mask, the planet model $m$ is simply null.

The flux posterior given the data is
\begin{align}
\mathcal{P}(F \vert D) &= \int_{S} \mathcal{P}(F ,S\vert D) \diff S\ ,\nonumber \\
&= \int_{S} \frac{\mathcal{P}(D\vert F ,S)\mathcal{P}(F ,S)}{\mathcal{P}(D)}\diff S\ ,\nonumber \\
&= \int_{S} \frac{\mathcal{P}(D\vert F ,S)\mathcal{P}(F )\mathcal{P}(S)}{\mathcal{P}(D)}\diff S,
\label{eq:orbitPoster}
\end{align}
where $\mathcal{P}(F)$ and $\mathcal{P}(S)$ are the priors for the point-source flux and position.

The likelihood is defined as,
\begin{equation}
\mathcal{P}(D = {\bm d}\vert F=f, S=s) = \frac{1}{\sqrt{2\pi\vert{\bm \Sigma}\vert}} \exp\left\lbrace-\frac{1}{2}({\bm d}-f{\bm m})^{\top}{\bm \Sigma}^{-1}({\bm d}-f{\bm m})\right\rbrace,
\label{eq:likelihood}
\end{equation}
which is the matrix version of a Gaussian likelihood seen in \autoref{eq:gausslike}.

For simplicity, we assume a uniform positive prior over flux,
\begin{equation}
\mathcal{P}(F=f) = \begin{cases} \alpha, & \mbox{if } f > 0\mbox{, with } \alpha \mbox{ constant} \\ 0, & \mbox{if } f \le 0 \end{cases} .
\end{equation}

The probability of finding the companion at any position in the orbit is proportional to the time spent around that position according to Kepler's laws. We can write:
\begin{equation}
\mathcal{P}(S) = \frac{1}{T v_{proj}}
\label{eq:sprior}
\end{equation}
where $T$ is the orbital period and $v_{proj}=v_{proj}(s)$ is the projected velocity of a companion at the position $s$ on the ellipse representing the projected orbit. $v_{proj}$ is constant in a face-on and circular orbit.
If $\mathcal{P}(S)$ is a delta function, we get the example from \Secref{sec:upperlimit}.

We have not yet specified the data term $d$ for the likelihood in \autoref{eq:likelihood}. It is possible to choose the final combined image, however the planet model might be poorly known when over- and self-subtraction from the speckle subtraction applies. Instead, we directly define the likelihood from the individual speckle subtracted images, which are for example defined by their exposure number and wavelength for an integral field spectrograph. 
The speckles are here subtracted using a principal component analysis (PCA)  based algorithm called Karhunen-Lo\'eve Image Projection \citep[KLIP,][]{Soummer2012}. KLIP consists in subtracting to each image its own projection on a subset containing $K$ elements of the principal components $z_{k}$. Defining the matrix $\boldsymbol{Z}_{K} = [ z_{1}, z_{2}, \dots, z_{K} ]^{\top}$, the speckle subtraction takes the form
\begin{equation}
{\bm i}_{\mathrm{sub}} = {\bm i} - \boldsymbol{Z}_{K}^{\top}\boldsymbol{Z}_{K} {\bm i},
\end{equation}
with $i$ the science image and ${\bm i}_{\mathrm{sub}}$ the speckle subtracted image.
Generally, the model of the signal is a function of the speckle subtraction algorithm used. When using a KLIP framework, point sources are distorted by the speckle subtraction. \cite{Pueyo2016} derived a linearized approximation of this distorted point spread function, which will be referred to as the forward model of the signal. Indeed, the existence of a faint point source in the data induces a perturbation on the principal components denoted $\Delta \boldsymbol{Z}_K$. We refer the reader to \cite{Pueyo2016} for the analytical expression of $\Delta \boldsymbol{Z}_K$, which is outside the scope of this paper. If ${\bm a}$ is the vectorized normalized planet signal in the science image, $f$ the planet flux and $n_i$ the associated noise containing the speckles, such that ${\bm i} = f{\bm a}+{\bm n_i}$, the normalized forward model then can be written
\begin{equation}
{\bm m} = {\bm a} - \boldsymbol{Z}_{K}^{\top} \boldsymbol{Z}_{K} {\bm a} - \left( \boldsymbol{Z}_{K}^{\top} \Delta \boldsymbol{Z}_K + \left(  \boldsymbol{Z}_{K}^{\top} \Delta \boldsymbol{Z}_K \right)^{\top} \right) \frac{{\bm i}}{f}.
\label{eq:fm}
\end{equation}
The linear approximation is valid when the planet signal is faint relative to the speckles (\textit{i.e.} f is small) and when there is little spatial overlap between the planet signal in the different reference images used in the principal components calculation.
The diagonal terms of the covariance matrix $\Sigma$ are directly estimated from the data as the empirical variance of each pixel.
With the assumption of Gaussian noise, the likelihood can be written directly as a function of the estimated flux $\tilde{f}_x$ and associated error bar $\sigma_x$ as defined in \cite{Ruffio2017} or \cite{Cantalloube2015}, with:
\begin{equation}
\tilde{f}_x = \frac{{\bm d}^{\top} {\bm \Sigma}^{-1} {\bm m}  }{{\bm m}^{\top} {\bm \Sigma}^{-1} {\bm m}},
\label{eq:maxlikf}
\end{equation}
and
\begin{equation}
\sigma_x^2 = \frac{1}{{\bm m}^{\top} {\bm \Sigma}^{-1} {\bm m}}.
\label{eq:sigf}
\end{equation}
The estimated flux $\tilde{f}_x$ is defined as the value maximizing the likelihood from \autoref{eq:likelihood}.
The terms $\Sigma$ and $\sigma_x$ should not be confused, the former characterizes the noise in the uncombined data and the latter represents the noise in the estimated flux map.
We can also write the theoretical matched filter S/N as
\begin{equation}
\mathcal{S}_x = \frac{{\bm d}^{\top} {\bm \Sigma}^{-1} {\bm m}  }{\sqrt{{\bm m}^{\top} {\bm \Sigma}^{-1} {\bm m}}}.
\label{eq:SNR}
\end{equation}
These quantities are the final products of matched-filter based data reduction \citep{Ruffio2017,Cantalloube2015}, which goal is to find the location of a known signal in noisy data. It is a maximum likelihood approach and consists in maximizing the S/N from \autoref{eq:SNR} as a function of the position of the planet. Substituting Equations \ref{eq:likelihood}, \ref{eq:fprior}, \ref{eq:sprior}, \ref{eq:maxlikf}, and \ref{eq:sigf} in \autoref{eq:orbitPoster}, we get the posterior
\begin{align}
\mathcal{P}(F=f \vert D={\bm d}) &\propto \mathcal{H}(f) \int_{s=s_i}^{s_f}
\exp\left\lbrace-\frac{1}{2}\left(f^2{\bm m}^{\top}{\bm \Sigma}^{-1}{\bm m}-2f{\bm d}^{\top}{\bm \Sigma}^{-1}{\bm m}\right)\right\rbrace
\frac{1}{T v_{proj}} \diff s \ ,\nonumber \\
&\propto \mathcal{H}(f) \int_{s=s_i}^{s_f}
\exp\left\lbrace-\frac{1}{2\sigma_s^2}\left(f^2-2f\tilde{f}_s\right)\right\rbrace
\frac{1}{T v_{proj}} \diff s .
\label{eq:postwithtildei}
\end{align}
We have used the fact that ${\bm d}^{\top}{\bm \Sigma}^{-1}{\bm d}$ does not depend on the position $s$ so we can factor it out of the integral as a proportionality constant.

In practice, the assumptions that were just made result in a biased estimate of the flux and the standard deviation. For example, the covariance ${\bm \Sigma}$ is not truly diagonal. The forward model is also not a perfect model of the planet, which underestimates the flux. As a consequence, we define the algorithm throughput as the ratio between the measured flux and the true flux of a point source and we denote it $\mu_x$.
A common practice to mitigate the standard deviation bias \citep{Ruffio2017,Cantalloube2015}, is to re-normalize the standard deviation such as to yield a S/N map with unit standard deviation ($\mathcal{S}_x/\eta_x \longrightarrow \mathcal{S}_x$, with the $\eta_x$ the standard deviation of the $\mathcal{S}_x$ map). The flux calibration is done with simulated planet injection and recovery ($\tilde{f}_x/\mu_x \longrightarrow \tilde{f}_x$), also known as algorithm throughput correction.
\autoref{eq:postwithtildei} now becomes:

\begin{align}
\mathcal{P}(F=f \vert D={\bm d}) &\propto \mathcal{H}(f) \int_{s=s_i}^{s_f}
\exp\left\lbrace-\frac{1}{2\eta_s^2\sigma_s^2}\left((\mu_s f)^2-2(\mu_x f)\tilde{f}_s\right)\right\rbrace
\frac{1}{T v_{proj}} \diff s.
\end{align}

\begin{figure}
\centering
\includegraphics[width=1\linewidth]{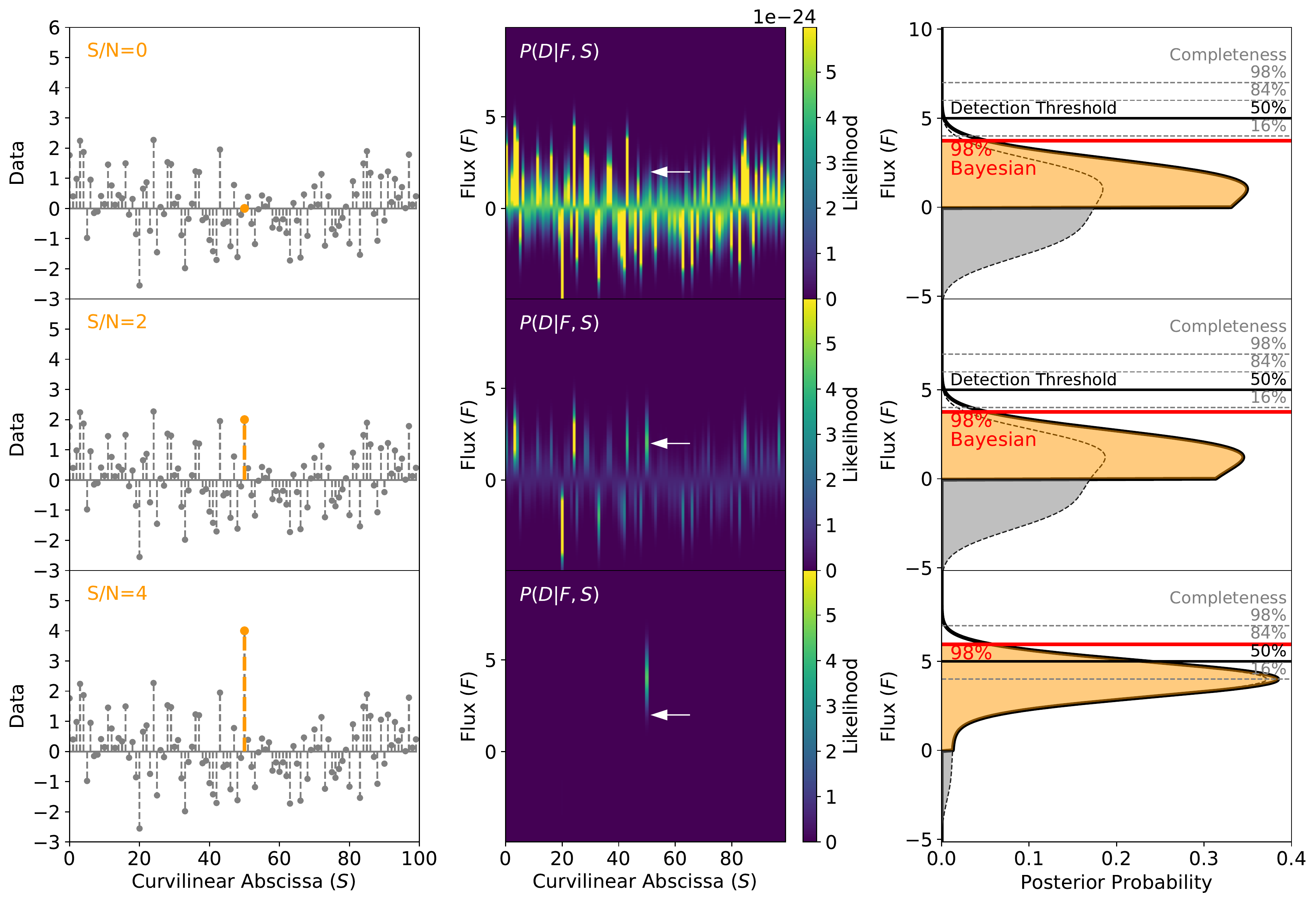}
\caption{Toy simulation illustrating Bayesian upper limits for planets on a known orbit but unknown location. The leftmost column features the observed flux as a function of the curvilinear abscissa of the orbit path. Each row corresponds to a given S/N of the pixel with curvilinear abscissa 50 highlighted by the vertical orange dashed line. The central column shows the joint likelihood as a function of flux and position of the planet in arbitrary units. The likelihood globally decreases as the signal increases because it becomes harder to explain with pure noise. The white arrow does not move in each plot and it highlights the location where the likelihood changes. The right column plots the flux posterior (black solid curve) of the planet compared to the completeness contours (grey horizontal lines) of a $5\sigma$ detection threshold (black horizontal lines). The dashed curve shows the posterior before the positive prior rejects the negative values of the flux (grey area). The Bayesian upper limit with $98\%$ cutoff probability is drawn as a horizontal solid red line. The posterior is marginalized over over $S$, which corresponds to a horizontal integral of the joint likelihood in the middle panels.  }
\label{fig:toyorbit}
\end{figure}

\autoref{fig:toyorbit} features a toy simulation of the approach assuming one hundred independent samples representing the pixel values along the orbit path, a unit standard deviation and a Dirac-like planet model (all the flux contained in one pixel). The value of the middle data point was fixed to a given S/N to show the effect of outliers on the upper limit. Similarly to \autoref{fig:posteriorKnownPos}, \autoref{fig:toyorbit} shows that the upper limit should truly be a function of the data ---in other words, a function of the realization of the noise---, which is not true of the detection threshold.

Note that the method is valid even when there is a strong (or weak) signal in the gap, which could be either a rare occurrence of the noise or an astrophysical object. This would not be true for a detection threshold based approach as the latter does not depend on the actual measurement.

\subsection{Effect of the size of the orbit}

\begin{figure}
\centering
\includegraphics[width=0.5\linewidth]{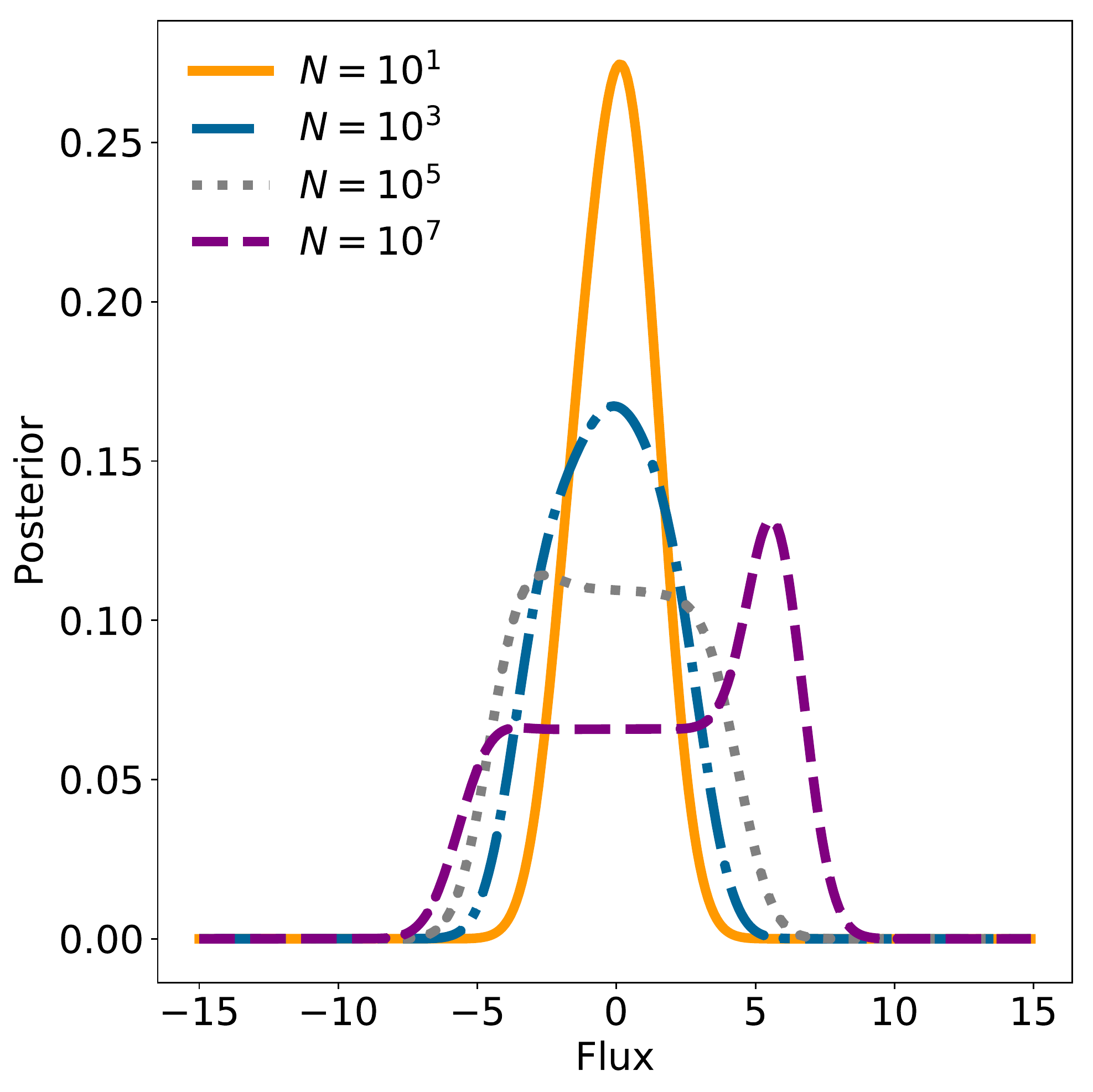}
\caption{flux posterior as a function of the number of elements using a simple simulation. The $N=10^7$ samples case highlights the effect of an outlier in the data.}
\label{fig:Nsamples}
\end{figure}

Intuitively, the upper limit is to the first order defined by the brightest signal in the data. As the number of elements increases, it becomes more likely to draw high S/N signals. Therefore, the larger the uncertainty on the location of the planet, the more realizations of the noise have to be considered and the poorer the upper limit will be. \autoref{fig:Nsamples} shows examples of posteriors when varying the number of samples in the \autoref{fig:toyorbit} simulation and assuming pure noise data with no real signal. Unless the data contains an outlier ---unlikely event considering the number of realizations---, the posterior flattens out as the number of samples increases.
At the limit of an infinite number of elements, the data loses any constraining power on the flux of the planet and the posterior becomes a constant.
When defining the upper limit from a detection threshold, this effect can be partially accounted for by modifying the detection threshold to yield the same false positive rate at any distance of the star \citep{JensenClem2018,Ruane2017}.

\subsection{Qualitative effect of non-Gaussianity and correlation}

\begin{figure}
\centering
\includegraphics[width=0.5\linewidth]{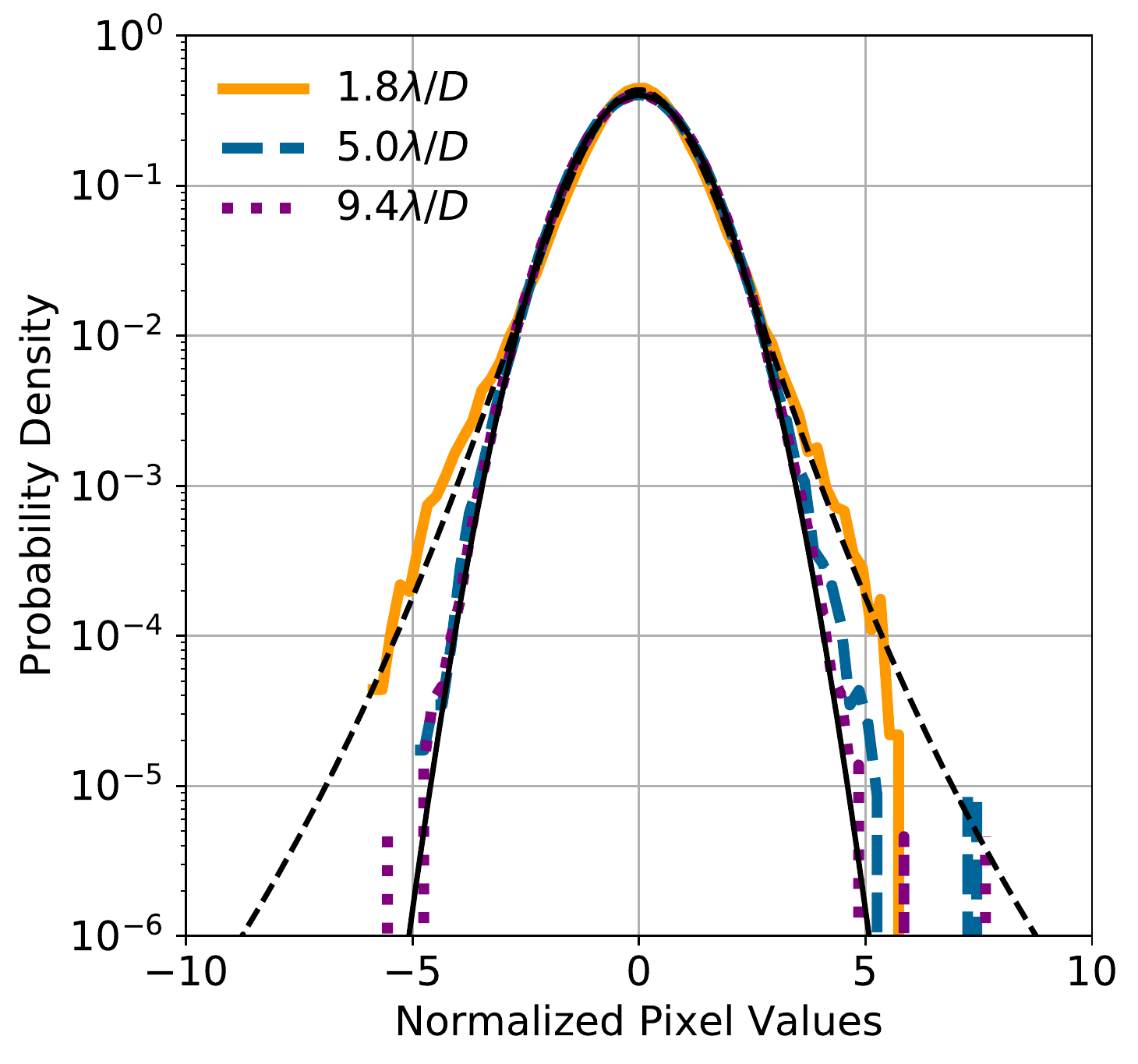}
\caption{Histogram of the speckle subtracted uncombined images for three separations for the NIRC2 observation of TW Hya. A resolution element is defined as $1.22\lambda/D\approx0.1''$ ($\lambda = 4\mu m$ and $D=10m$).}
\label{fig:klippdfvssep}
\end{figure}

\begin{figure}
\centering
\includegraphics[width=1\linewidth]{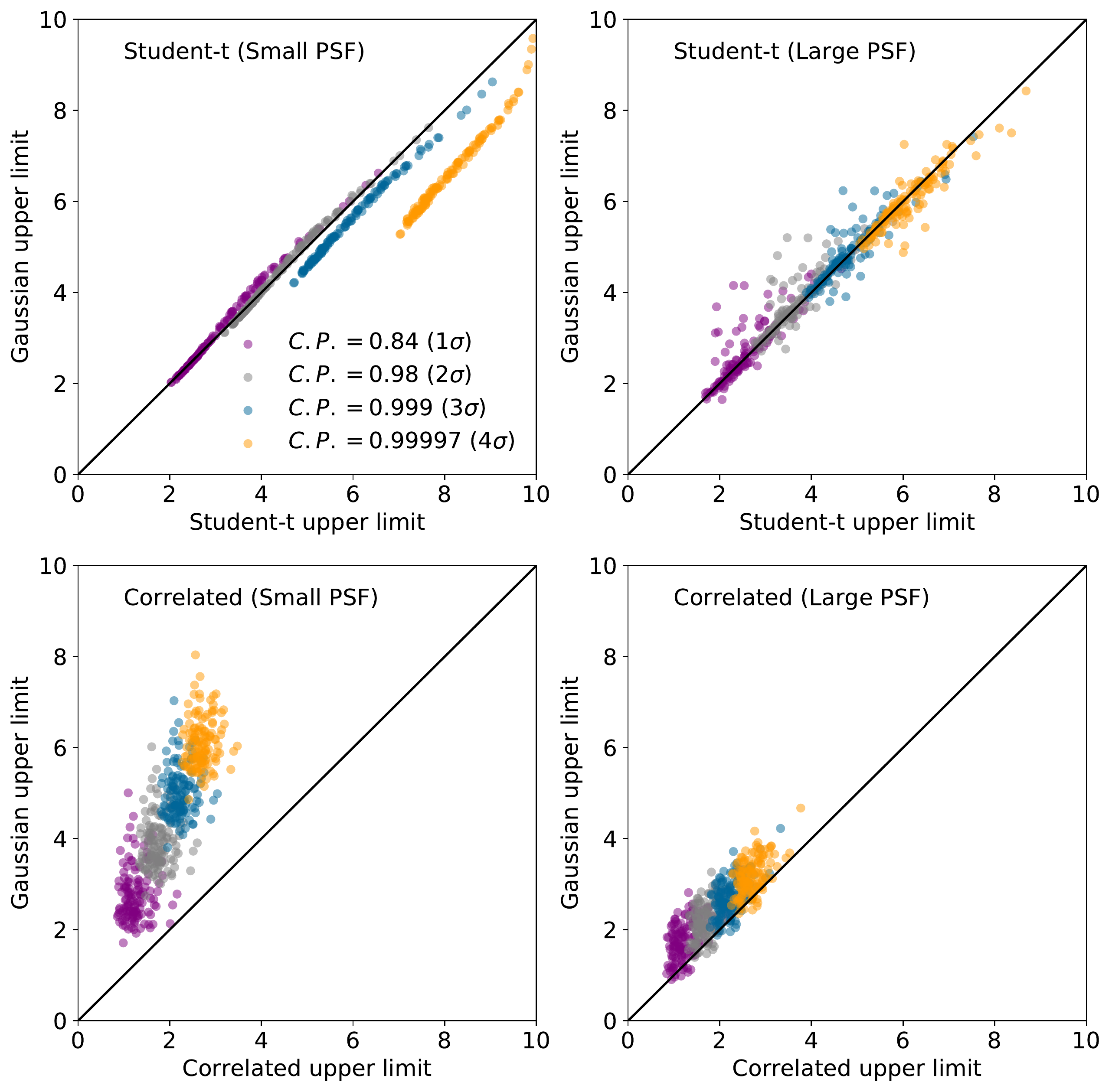}
\caption{Comparison of the upper limits derived from the true properties (x-axis) of the noise with the upper limits derived with the assumption of Gaussian and independent noise (y-axis). The color corresponds to different cutoff probability (CP). The diagonal represents a correct estimation of the upper limit despite the approximation that is made. Points below the diagonal show that the upper limits has been underestimated leading to over-confident constraints. Points above the diagonal represent over-estimated upper limits leading to conservative results.
Units are arbitrary.}
\label{fig:upperlimitsComparisons}
\end{figure}

We have assumed that the noise was Gaussian and un-correlated. \autoref{fig:klippdfvssep} shows the histogram of the pixel values in uncombined speckle-subtracted images for three different separations. While the distribution approximate a Gaussian for the largest two separations, it has a very large tail in the inner-most case.
In this section, we simulate the effects of the correlation and the non Gaussianity on the final upper limit in a simplified example. We assume a discrete orbit made of hundred elements similar to \autoref{fig:toyorbit}, containing pure noise. In order to explore the effect of the size of the planet with respect to non Gaussian noise and non diagonal correlation matrix, we consider two models: a small Dirac-like point spread function (PSF) and a large PSF five-elements wide. In \autoref{fig:upperlimitsComparisons}, we estimate the error made on the upper limit when we erroneously assume that the noise is Gaussian or independent. In order to do so, we compare the upper limits derived from the true properties of the noise with the upper limits derived with the assumption of Gaussian and independent noise. 
The non-Gaussianity is first tested using a Student-t distribution with ten degrees of freedom and normalized to unit standard deviation (top panels of \autoref{fig:upperlimitsComparisons}), which was a good fit to the pixel distribution of the inner-most annulus in \autoref{fig:klippdfvssep}. In this context, the Student-t distribution is used for its wider tails compared to a Gaussian distribution, but not because of its relation to small sample statistics. Then, we try a correlated Gaussian noise with a circulant covariance matrix\footnote{A circulant matrix is a matrix for which each row vector is shifted by one element to the right relative to the preceding row vector. It is a special case of Toeplitz matrix for which the diagonals are constant. The circulant matrix is here used to express the periodicity of the projected orbit.} (bottom panels of \autoref{fig:upperlimitsComparisons}). The correlation profile is defined as a Gaussian with a standard deviation equal to two elements. 
The upper limits are calculated in a similar fashion as for \autoref{fig:toyorbit} by replacing the likelihood from \autoref{eq:orbitPoster} to include a non-diagonal covariance matrix or to use a student-t distribution. The student-t joint likelihood is computed as the product of the individual likelihood for each pixel, which is granted by the independence of the noise.
Despite the simplicity of these simulations, we can draw some general principles from it.
First, it is necessary to account for the non-Gaussian tail of the noise only when considering high cutoff probabilities (e.g. $>0.999$, which is equivalent to $3\sigma$).
It is therefore good practice to quote upper-limits derived from reasonable cutoff probabilities if the distribution of the noise is poorly known.
In this crude simulation, the non-Gaussian noise is uncorrelated, which means that the effect is mitigated when the planet PSF is large. Indeed, the noise becomes more Gaussian when combining several pixels together because of the central limit theorem. In practice, the non-Gaussian noise comes from the correlated speckle noise with a correlation length equal to the PSF size, which means that a larger PSF will not help. However, the noise will be made more Gaussian thanks to the observing strategies ---Angular Differential Imaging (ADI) \citep{Marois2006} and Spectral Differential Imaging (SDI) \citep{Marois2000,Sparks2002}--- where the quasi-static speckle are subtracted and the displacement of the planet relative to them is used.

Secondly, when neglected, correlated noise can create the illusion of a signal, which results in inflated upper limits. However, overestimated upper limits are a conservative choice, which makes it acceptable. This effect is partially corrected when we calibrate the standard deviation as discussed at the end of \Secref{sec:knownOrbitdef}. Note that a large PSF here again mitigates the effects of the correlation. Increasing the correlation length and the PSF size in tandem is equivalent to reducing the number of independent realizations of the noise.

\subsection{Constraining the mass of a planet in the TW Hya protoplanetary disk}

We apply the previous framework to the Keck-NIRC2 observations of TW Hya at \textit{L'} presented in \cite{Ruane2017}. TW Hya features a proto-planetary disk in which gaps have been detected \citep{vanBoekel2017,Andrews2016,Debes2016,Rapson2015,Akiyama2015,Weinberger2002}. A possible explanation for the gaps could be the presence of accreting planets carving them \citep{Dong2017}. The goal is to use the high contrast observation of the system to set an upper limit on the mass of the hypothetical planets, which can be compared to the predictions of theoretical models of planet formation.

We will compare the different definitions of upper limit. 
As a reminder, our Bayesian upper limits are defined from the flux posterior and the probability of the true planet flux (or mass) to be smaller than the limit, which we called cutoff probability.
The frequentist definition of upper limit relies on the detection threshold and its associated completeness. A detection occurs when the measured flux of a point source falls above the detection threshold. The completeness is the probability that a planet of a given flux (or mass) is detected. The completeness is therefore always $50\%$ for planets with a true flux equal to the detection threshold. Otherwise, it will, for example, be equal to $16\%$ if the true flux of the planet is $1\sigma$ below the detection threshold and $84\%$ if it is $1\sigma$ above the detection threshold.
To summarize, the frequentist upper limit makes a statement about the detectability of a planet, while the Bayesian upper limit makes a statement about the probability of the planet flux given the data.

\begin{figure}
\gridline{\fig{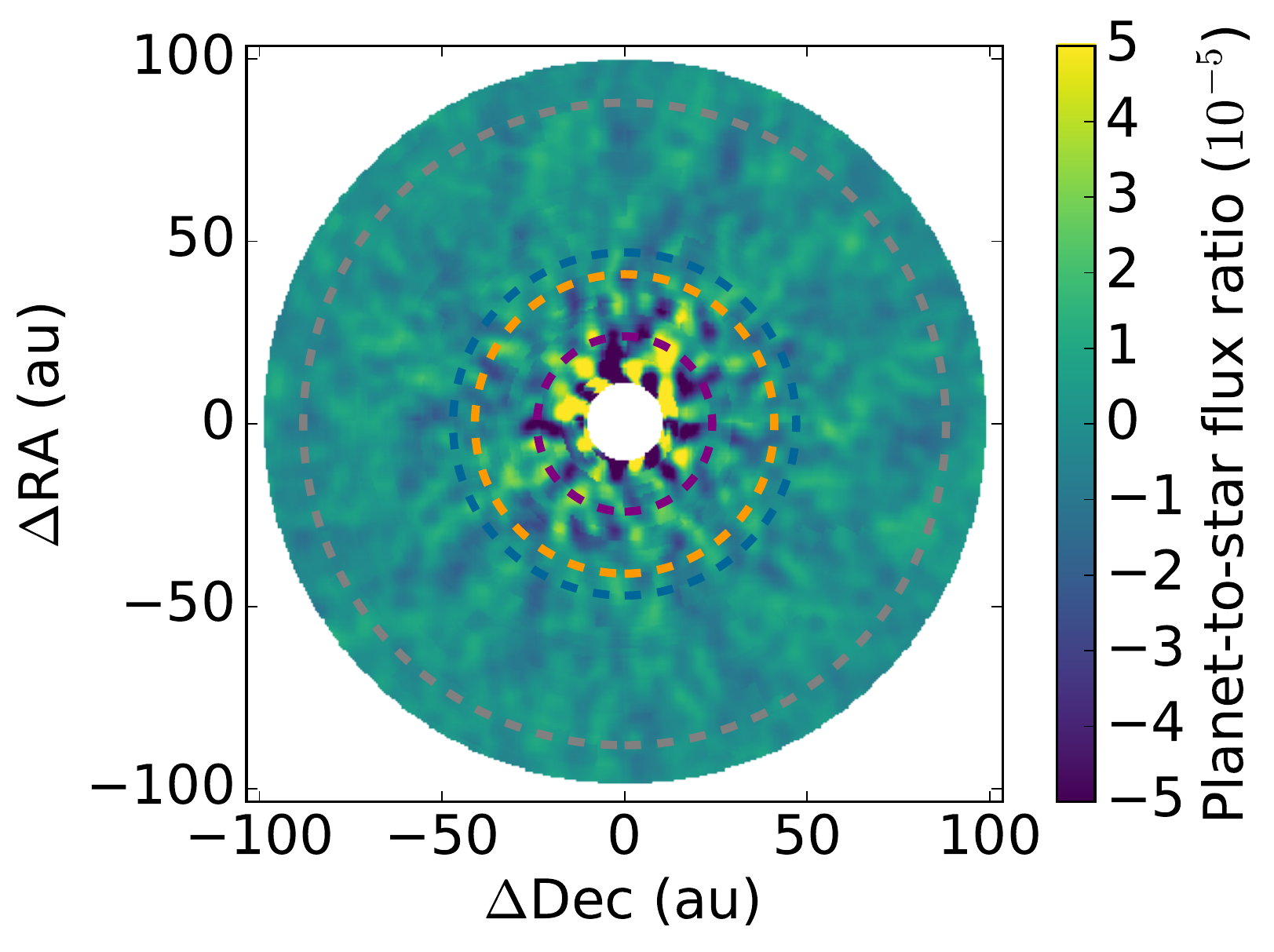}{0.5\textwidth}{(a) Planet-to-star flux ratio ($\tilde{f}_x$)}
		\fig{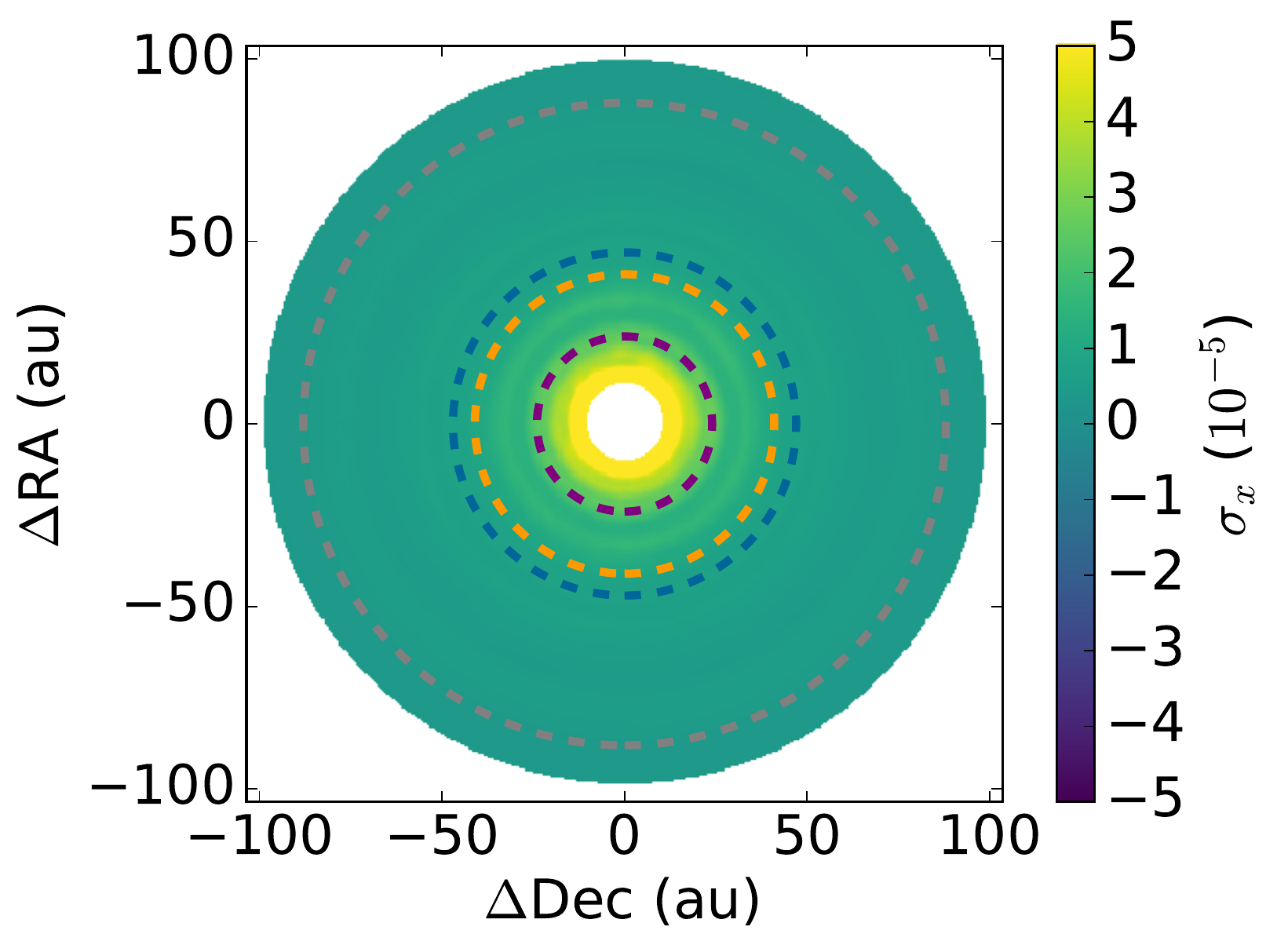}{0.5\textwidth}{(b) Flux standard deviation ($\sigma_x$)}
          }
\caption{(a) Planet-to-star flux ratio of TW Hya and (b) its corresponding standard deviation maps. The maps were calculated according to \autoref{eq:maxlikf} and \autoref{eq:sigf} using a forward model matched filter described in \cite{Ruffio2017}. The dashed lines represent the known gaps in the protoplanetary disk at $24$,$41$,$47$, and $88$ au.  \label{fig:TWHyacontsig}}
\end{figure}

\begin{figure}
\gridline{\fig{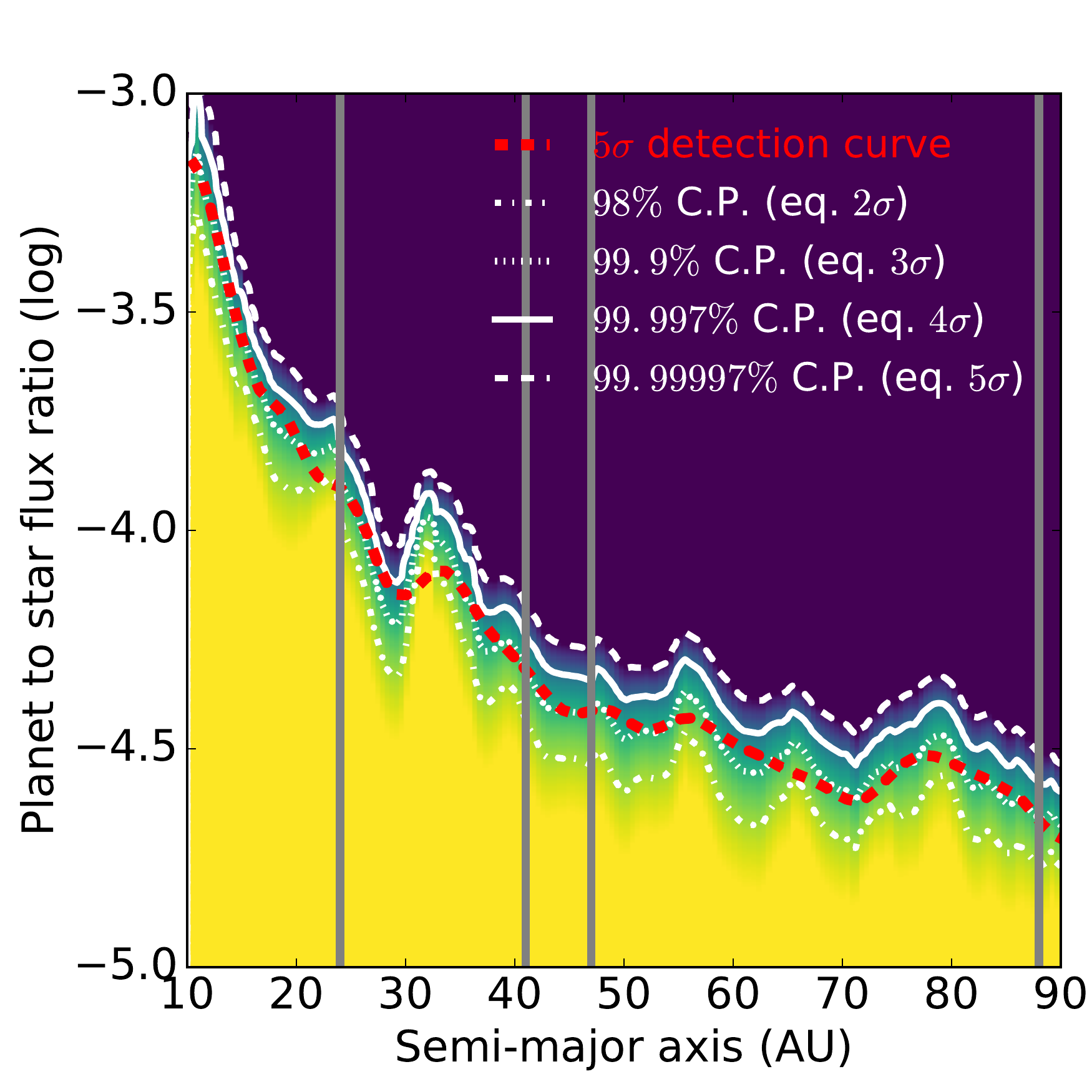}{0.5\textwidth}{(a) Flux ratio}
		\fig{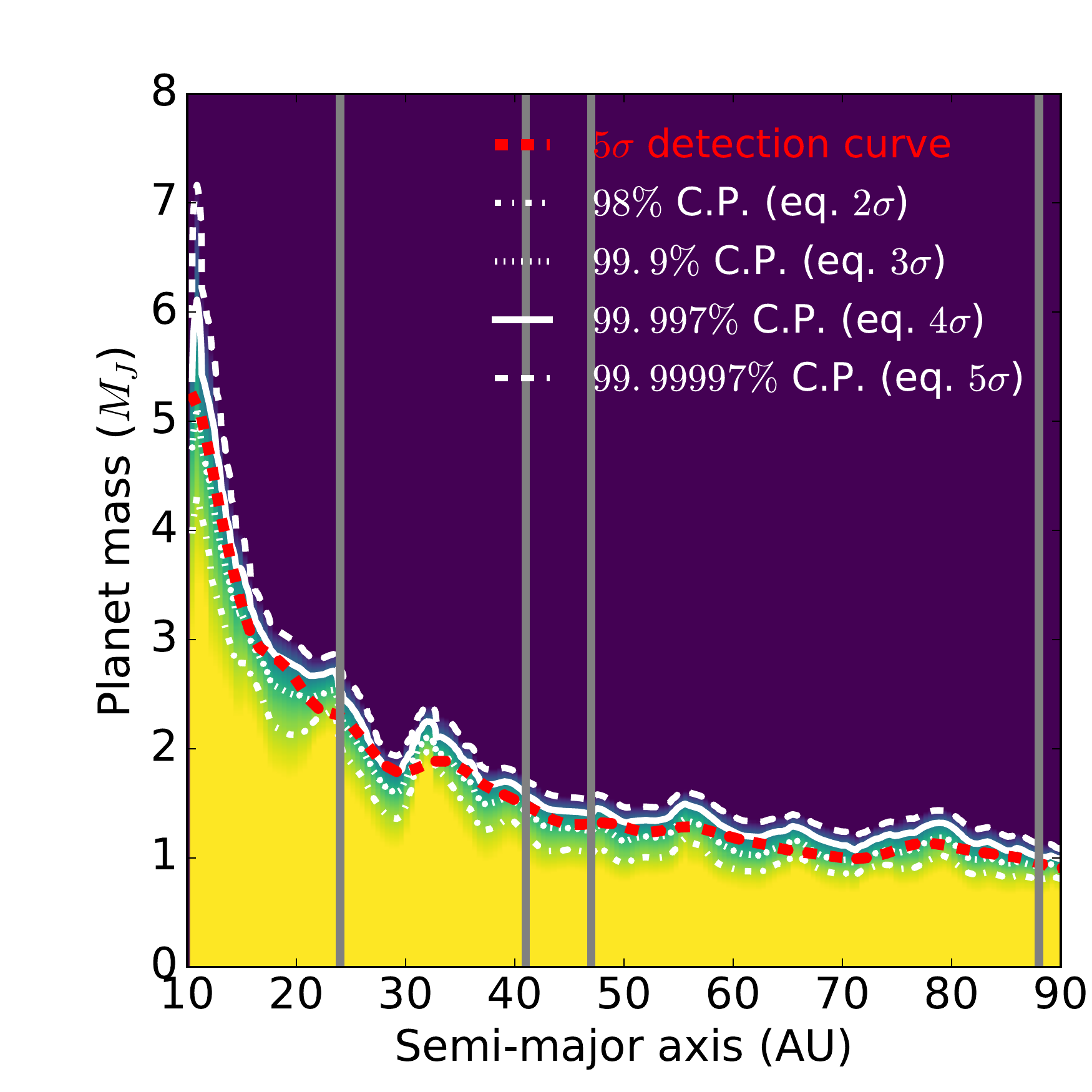}{0.5\textwidth}{(b) Mass}
          }
\caption{Cumulative distribution of the planet-to-star flux ratio (respectively mass) posterior for TW Hya as a function of the semi-major axis of the putative planet to the star. The Bayesian upper limit contours are drawn for different probabilities and can be compared to the conventional $5\sigma$ detection threshold. The equivalent significance of each cutoff probability is also written in terms of sigmas in the legend. The location of the gaps are marked with a vertical gray solid line. The flux ratio to mass conversion was performed using the AMES-Cond model and a $10$Myr old star.  \label{fig:TWHyaCDFpost}}
\end{figure}

The units of the planet flux has not yet been defined and will depend on the normalization of the planet model. In the following, we will express the flux as the planet-to-star flux ratio.
\autoref{fig:TWHyacontsig} shows the planet-to-star flux ratio of TW Hya and its corresponding standard deviation maps, which are used in the calculation of the likelihood in \autoref{eq:postwithtildei}.
\autoref{fig:TWHyaCDFpost} shows the resulting planet-to-star flux ratio and the mass upper limits for different cutoff probabilities as a function of the planet semi-major axis (white lines). It also features the $5\sigma$ detection threshold as a function of separation (dashed red line). The location of the four gaps in TW Hya proto-planetary disk are marked with grey lines.
The planet to star flux ratio to mass conversion was performed using the AMES-Cond model \citep{Baraffe2003,Allard2012} and an age of $10$Myr for the star \citep{Bell2015}, which only accounts for photospheric emissions and neglect possible effects of accretion. We assume a uniform positive prior in flux or mass for the calculation of the flux or mass posterior respectively. It means that the priors in both cases, mass or flux, are not equivalent. We could use a more informed mass prior based on observational results, but it is still poorly constrained for giant planets at large separation and the choice of a constant is conservative. 

\begin{figure}
\centering
\includegraphics[width=1\linewidth]{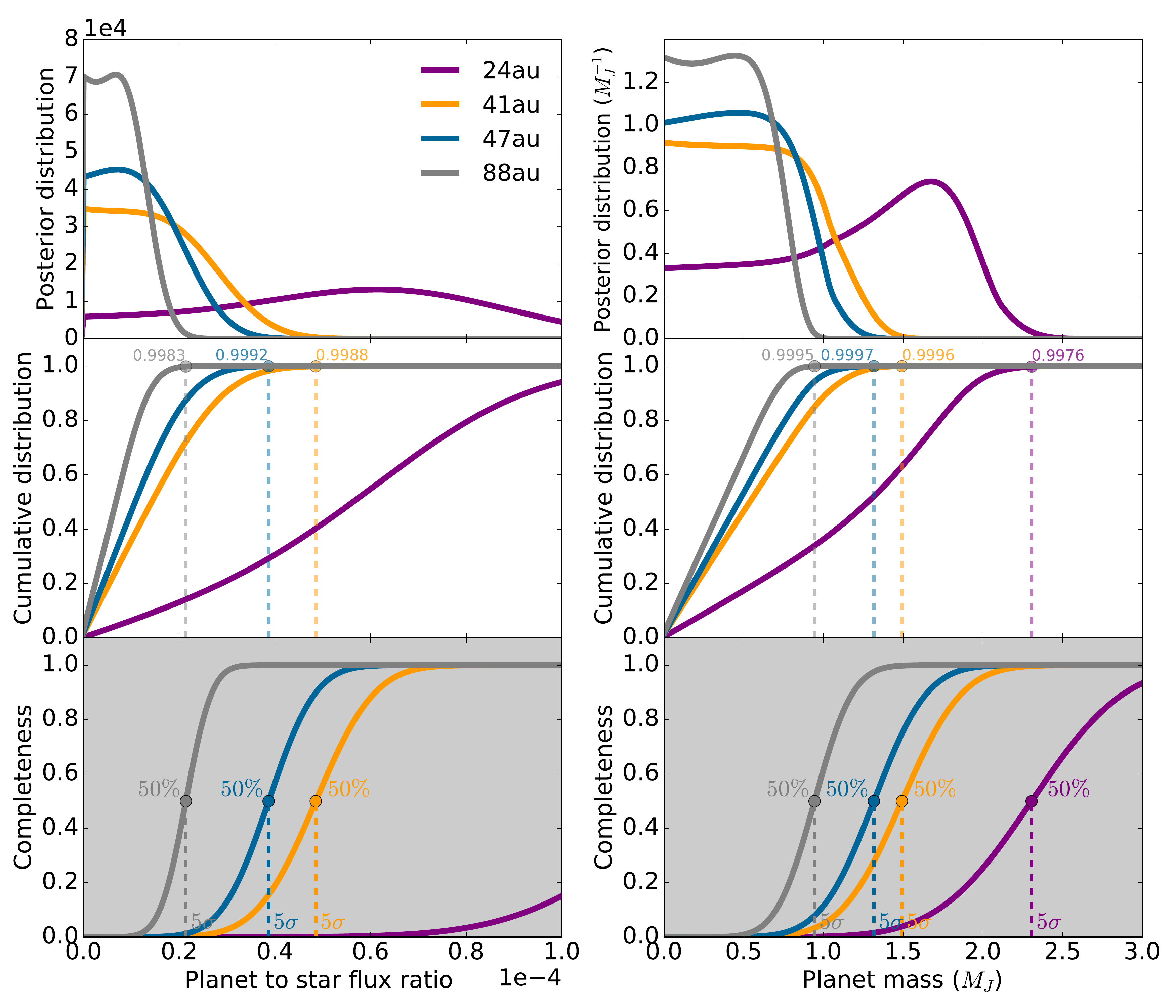}
\caption{
Posterior, cumulative distribution assuming a positive prior and completeness for each gap of \autoref{fig:TWHyaCDFpost}. The first column features the distributions as a function of the planet-to-star flux ratio while the second column uses the planet mass. The completeness is calculated for a $5\sigma$ detection threshold, which itself corresponds to a $50\%$ completeness. The value of the cumulative distribution of the posterior (\textit{ie} the cutoff probability) corresponding to the detection threshold is indicated above the curve for each gap. \label{fig:cutsgaps} }
\end{figure}

\autoref{fig:TWHyaCDFpost} does not allow for an easy comparison of the different upper limits. \autoref{fig:cutsgaps} features the posterior and its cumulative distribution as well as the completeness for each gap, which are in substance vertical cuts through \autoref{fig:TWHyaCDFpost}. For example, on one hand, the detection threshold corresponds to a $\approx1M_\mathrm{J}$ planet at the 88 au gap (grey lines). A possible definition of mass upper limit would be the mass of a planet that would be detected $95\%$ of the time, which is $\approx1.3M_\mathrm{J}$ for that gap. On the other hand, the value of the cumulative distribution of the posterior at the detection threshold is already 0.9995 ($\approx 3\sigma$), which already highlight a comfortable degree of confidence in the fact that the mass of the planet is lower. In hindsight, it seems unnecessary to look for such a high completeness in this case. It is also common practice to quote the detection threshold itself as the upper limit. This approach can lead to confusion because the $5\sigma$ of the detection threshold can be easily mistaken for a cutoff probability, when it is really associated to a false positive rate. A $5\sigma$ detection threshold is not equivalent to a Bayesian upper limit with a ``$5\sigma$'' cutoff probability, but truly $\approx3\sigma$ in the specific example. An upper limit should always be accompanied by a statement about its probability, which a sole detection threshold is lacking.

We have already mentioned other caveats coming from a detection-based upper limit. 
First, the definition of the threshold is somewhat arbitrary. The field has widely adopted a $5\sigma$ threshold but it yields many false positives in practice due to the non-Gaussianity of the noise. If we chose a larger threshold to mitigate this issue, should the upper limit change or remain the same?
Secondly, a detection threshold only indirectly depends on the data through the sample standard deviation, while the Bayesian upper limit  fully expresses the information contained in the data. The latter will be highly sensitive to outliers. This dependence to the data is illustrated in \autoref{fig:TWHyaCDFpost}, where the Bayesian upper limit varies strongly as a function of the semi-major axis while the detection threshold is smoother.
Finally and related to the previous point, an advantage of the Bayesian upper limit is that it is valid regardless of the strength of the signal (detection or non-detection), while the detection threshold framework requires the absence of outliers. Besides, the Bayesian approach is less sensitive to the assumed noise distribution than the frequentist approach due to the dominant impact of the strongest signal in the data.

In \cite{Ruane2017}, the upper limit is calculated for a $95\%$ completeness and a detection threshold that is defined to yield 0.01 false positives within $1''$ of the host star. The threshold varies from $8.1\sigma$ to $4.5\sigma$ with increasing separation to the star accounting for both the larger area available at larger separation and small sample statistics using a Student-t distribution \citep{Mawet2014}. The most conservative upper limits for the mass of a companion around TW Hya assume an age of $10$ Myr and the AMES-Cond model. The reported upper limits for each gap in the system are: $2.3\,M_{\mathrm{J}}$ at $24$ au, $1.6\,M_{\mathrm{J}}$ at $41$ au, $1.5\,M_{\mathrm{J}}$ at $47$ au, and $1.2\,M_{\mathrm{J}}$ at $88$ au.
Using a Bayesian analysis and $0.999$ cutoff probability ($3\sigma$), this work finds mass upper limits equal to $2.4\,M_{\mathrm{J}}$ at $24$ au, $1.5\,M_{\mathrm{J}}$ at $41$ au, $1.3\,M_{\mathrm{J}}$ at $47$ au, and $0.9\,M_{\mathrm{J}}$ at $88$ au. Note that this work uses a different reduction algorithm compared to \cite{Ruane2017}, which has not been optimized and might explain the limited gains. \autoref{fig:cutsgaps} should be used for a fair comparison of the frequentist and Bayesian approach.
The upper limits for the planet-to-star flux ratio are $1.4\times10^{-4}$ at $24$ au, $4.9\times10^{-5}$ at $41$ au, $3.4\times10^{-5}$ at $47$ au, and $2.2\times10^{-5}$ at $88$ au. The corresponding absolute magnitudes are respectively 12.9, 14.0, 14.3, and 14.9, where we have assumed a TW Hya distance of $60.1\pm0.15$ pc \citep{Gaia2018,Gaia2016} and a 7.01 apparent magnitude at Wise W1 band (proxy for  Keck-NIRC2 L' band)  \citep{Wright2010}.

\begin{figure}
\centering
\includegraphics[width=1\linewidth]{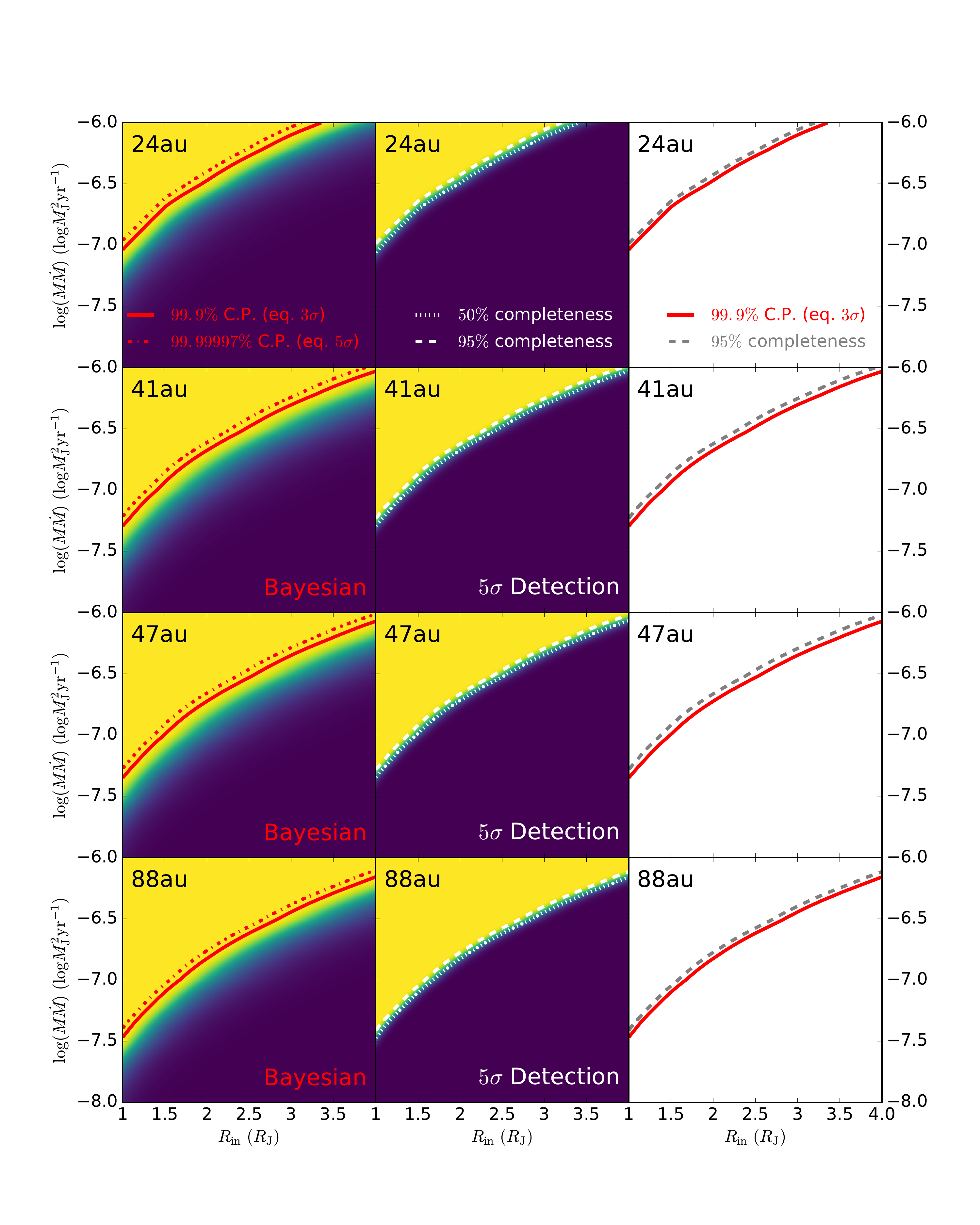}
\caption{
Cutoff probabilities and detection completeness as a function of $(M\dot{M}, R_{\mathrm{in}})$ for each gap (rows) in TW Hya protoplanetry disks. The L'-band absolute magnitudes for each value of $(M\dot{M}, R_{\mathrm{in}})$  were linearly interpolated from the table in \cite{Zhu2015}. The left images show the cutoff probability, which is the value of the cumulative distribution of the posteriors in \autoref{fig:cutsgaps} at the flux of the planet. The middle images show the corresponding completeness for a $5\sigma$ detection threshold. The right plots compare the $99.9\%$ cutoff probability with the $95\%$ completeness contours.
 \label{fig:MMdot} }
\end{figure}

The previous mass upper limits do not consider possible accretion of the planet, which is most likely to occur for proto-planetary disks like TW Hya. The absolute magnitude of a circumplanetary disk is a function of the product of the planet mass with the accretion rate, $M\dot{M}$, and the inner radius of the circumplanetary disk, $R_{\mathrm{in}}$ \citep{Zhu2015}. For small enough planets, which is most often the case, the intrinsic flux of the planet can be neglected as the accretion appears much brighter. We can compute the probability corresponding to each set of parameters $(M\dot{M}, R_{\mathrm{in}})$ using \cite{Zhu2015} model predictions for the circumplanetary disk absolute magnitude in L band and the posterior distribution of \autoref{fig:cutsgaps}. We compare the map of cutoff probabilities and completeness for each gap in \autoref{fig:MMdot}. On a logarithmic scale, the $99.9\%$ Bayesian cutoff probability leads to marginally better constraints on $(M\dot{M}, R_{\mathrm{in}})$ than the $5\sigma$ completeness based approach. However, we argue that it is easier to statistically interpret the Bayesian constraints. \cite{Dong2017} argues that hypothetical gap opening planets should have mass of the order of $0.1M_{\mathrm{J}}$ in the $20$au and $80$au gaps. The thermal emission of such planets would be invisible at our current sensitivity and therefore can be neglected. Therefore, assuming a $0.1M_{\mathrm{J}}$ planet, we can set constraints on the accretion rate $\dot{M}$ by adding one unit to the y-logarithmic-scale in \autoref{fig:MMdot}. As a conclusion, given the data, we are $99.9\%$ confident that the accretion rate of a $0.1M_{\mathrm{J}}$ planet in the TW Hya gaps, if it exists, is below $9.3\times10^{-7}\,M_{\mathrm{J}}\mathrm{yr}^{-1}$ at $24$ au, $5.0\times10^{-7}\,M_{\mathrm{J}}\mathrm{yr}^{-1}$ at $41$ au, $4.5\times10^{-7}\,M_{\mathrm{J}}\mathrm{yr}^{-1}$ at $47$ au, and $3.4\times10^{-7}\,M_{\mathrm{J}}\mathrm{yr}^{-1}$ at $88$ au.
 
\clearpage

\section{Combining radial velocity and direct imaging observations}
\label{sec:RVIM}

Combining Radial Velocity (RV) data with direct imaging is another very promising avenue for constraining masses of non-transiting wide-orbit planets detected with Doppler measurements. An example of application of this method can be found in the case of \epseri\ in Mawet et al. (2018, submitted to AJ) for which this section describes the theoretical concepts.

Radial velocity only provides a lower limit on the mass of the planet due to the $M\sin(i)$ mass-inclination degeneracy. Direct imaging non detection can provide a mass upper bound and therefore reject the lower inclinations. 
We define $D_{RV}$ as the time series of radial velocities and $D_{DI}$ as the direct imaging observation. The model parameters $\Theta$ to be inferred include the orbital elements and the mass of the planet as well as the star. We also define $\Theta'$ such that $\Theta=\left\lbrace M_{p},\Omega,\Theta'\right\rbrace$ with $M_{p}$ the mass of the planet and $\Omega$ the position angle or longitude of ascending node.

$D_{RV}$ and $D_{DI}$ are independent so the posterior can be written as,
\begin{align}
\mathcal{P}(\Theta \vert D_{RV},D_{DI}) &= \frac{\mathcal{P}(D_{RV},D_{DI} \vert \Theta) \mathcal{P}(\Theta)}{\mathcal{P}(D_{RV},D_{DI})}, \nonumber\\
&= \frac{\mathcal{P}(D_{RV} \vert \Theta)\mathcal{P}(D_{DI} \vert \Theta) \mathcal{P}(\Theta)}{\mathcal{P}(D_{RV},D_{DI})}.
\end{align}

The radial velocity log-likelihood can be written as \citep{Howard2014}
\begin{equation}
\log{\mathcal{P}(D_{RV}\vert \Theta)} = -\sum_i{\left[\frac{(v_i-v_m(t_i))^2}{2(\sigma_i^2 + \sigma_j^2)} + \log{\sqrt{2\pi(\sigma_i^2+\sigma_j^2)}}\right]},
    \label{eq:rvlikelihood}
\end{equation}
where $v_i$ are the measured radial velocities at the times $t_i$ and  $v_m(t_i)$ are the corresponding projected Keplerian velocities. The standard deviations $\sigma_i$ and $\sigma_j$ are respectively the internal uncertainty for each measurement and the instrument-specific jitter term. 

As it was shown in \Secref{sec:knownOrbitdef}, the direct imaging log-likelihood can be written as
\begin{equation}
    \log{\mathcal{P}(D_{DI}\vert F=f, X=x)} \propto -\frac{1}{2\sigma_x^2}\left(f^2-2f\tilde{f}_x\right).
    \label{eq:dilikelihood}
\end{equation}
The planet flux $f$ is a function of the planet mass and stellar age and the position $x$ is determined by the orbital parameters.

The posterior on $\Theta$ can be inferred using a Markov Chain Monte Carlo (MCMC).
Keeping both spatial dimensions in the direct imaging likelihood, like in Mawet et al. (2018, submitted to AJ), requires a much longer Markov chain to converge.
It is possible to make the problem more tractable by marginalizing the problem over the position angle of the planet with minimal loss of information. Indeed, the radial velocity cannot constrain this parameter and the image has little valuable information about the position angle in the absence of an obvious outlier.
\begin{align}
  \mathcal{P}( M_p,\Theta' \vert D_{RV},D_{DI}) &=  \int_{\omega} \mathcal{P}(\Theta \vert D_{RV},D_{DI}) \nonumber\\
&= \int_{\omega}  \frac{\mathcal{P}(D_{RV} \vert \Theta)\mathcal{P}(D_{DI} \vert \Theta) \mathcal{P}(\Theta)}{\mathcal{P}(D_{RV},D_{DI})} \nonumber\\
&\propto \mathcal{P}(D_{RV} \vert M_p,\Theta')\mathcal{P}(\Theta'') \mathcal{P}(M_{p}) \left[ \int_{\omega}  \mathcal{P}(D_{DI} \vert F, r,\omega)\mathcal{P}(\omega) \right] \label{eq:marginalizedRVDI}
\end{align}
with $\mathcal{P}(\Omega)=1/2\pi$ the prior for the position angle and $r,\omega$ are the cylindrical coordinates of the planet.

\autoref{eq:marginalizedRVDI} shows that the radial velocity likelihood can be factored out of the integral, which leaves an integral over the sole direct imaging likelihood. The calculation of the integral is then equivalent to \Secref{sec:knownOrbitdef} with a circular face-on orbit.
Note that it is always possible to derive an upper limit, even when the existence of a planet is still in doubt. For example, one could use this framework to constrain masses for radial velocity trends.

\section{Conclusion}
\label{sec:conclusion}

In this work we have addressed the differences between frequentist and Bayesian definitions of planet flux upper limits in the context of exoplanet direct imaging.
The frequentist upper limit makes a statement about the detectability of a planet, while the Bayesian upper limit is about the probability of a given planet flux. While upper limits are often thought of in Bayesian way, they are mostly quoted as detection threshold in our field. This makes the interpretation of detection based upper limit more challenging.
The detection threshold, or contrast curve, is somewhat arbitrary and only a property of the noise ($\sigma$), which means it is not an optimal use of the data. 

Our goal is to provide a conceptual framework for the analysis of direct imaging data as well as informative examples rather than an explicit analysis recipe or formalism.
Clearly defining a problem and its statistical representation before any calculation is extremely important. Our conceptual framework can also be applied to other cases such as estimating planet occurrence rates, or combining multiple measured quantities such as relative motion. Here we have illustrated three typical cases: 
\begin{itemize}
\item deriving an upper limit on planet flux when the location of the planet is known, which happens when e.g. the planet is robustly detected in a set of filters but not in others. The set of observations can be used to inform the astrophysical nature of a candidate (planet, brown dwarf, star, galaxies\dots) using Bayesian model comparison.
\item constraining the mass of a hypothetical planet carving a gap in a protoplanetary disk \citep{Ruane2017}. We showed that the data contains more information and is typically more constraining than the sole detection threshold suggests. Illustrating our method on the TW Hya system, there is a $99.9\%$ probability given the data that the mass of hypothetical non-accreting planets in the gaps are below $2.4\,M_{\mathrm{J}}$ at $24$ au, $1.5\,M_{\mathrm{J}}$ at $41$ au, $1.3\,M_{\mathrm{J}}$ at $47$ au, and $0.9\,M_{\mathrm{J}}$ at $88$ au. With the same probability, the accretion rate of a $0.1M_{\mathrm{J}}$ planet with a $1 R_{\mathrm{J}}$ circumplanetary disk inner radius, if they exist, is below $9.3\times10^{-7}\,M_{\mathrm{J}}\mathrm{yr}^{-1}$ at $24$ au, $5.0\times10^{-7}\,M_{\mathrm{J}}\mathrm{yr}^{-1}$ at $41$ au, $4.5\times10^{-7}\,M_{\mathrm{J}}\mathrm{yr}^{-1}$ at $47$ au, and $3.4\times10^{-7}\,M_{\mathrm{J}}\mathrm{yr}^{-1}$ at $88$ au.
\item We also introduced the problem of combining the radial velocity and direct imaging measurement, where a joint Bayesian likelihood brings out the power of the two methods and can be used to infer the mass and orbital parameters of a planet (see Mawet et al., 2018, submitted to AJ).
\end{itemize}

\acknowledgments
{\bf Acknowledgements:} The research was supported by grants from NSF, including AST-1411868 (J.-B.R., B.M.) and AST-1518332 (R.J.D.R.).
Support was provided by grants from NASA, including NNX14AJ80G (B.M., J.-B.R.), NNX15AD95G (R.J.D.R.) and NNX15AC89G (R.J.D.R.). This work benefited from NASA’s Nexus for Exoplanet System Science (NExSS) research coordination network sponsored by NASA’s Science Mission Directorate.

\vspace{5mm}
\facilities{KeckII(NIRC2)}

\software{pyKLIP\footnote{Documentation available at \url{http://pyklip.readthedocs.io/en/latest/}.} \citep{Wang2015}, astropy\footnote{\url{http://www.astropy.org}} \citep{Astropy2013}, Matplotlib\footnote{\url{https://matplotlib.org}} \citep{Hunter2007}}

\newpage
\bibliographystyle{aasjournal}
\bibliography{bibliography}

\end{document}